\documentclass[a4paper,11pt]{article}
\usepackage{jinstpub}
\usepackage{lineno}

\usepackage{graphicx}
\usepackage{tabularx}
\usepackage{threeparttable}
\usepackage{subcaption}
\usepackage{siunitx}
\usepackage{microtype}
\usepackage{booktabs}
\usepackage{multirow}

\usepackage{tikz}
\usetikzlibrary{arrows.meta,positioning,shapes.geometric}

\usepackage{textcomp}

\usepackage[section]{placeins}
\usepackage{float}

\DeclareSIUnit{\photon}{ph}
\DeclareSIUnit{\MeV}{MeV}
\DeclareSIUnit{\photonperMeV}{\photon\per\MeV}

\title{Absolute scintillator light yield correction for SiPIN readout via Transfer Matrix Method and Geant4 optical simulation}

\author[b]{Ge Ma}
\author[a,1]{Zhiyang Yuan\note{Corresponding author.}}
\author[a]{Chencheng Feng}
\author[b]{Zirui Yang}
\author[a]{Zhenwei Yang}
\author[b]{Ming Zeng}

\affiliation[a]{School of Physics, State Key Laboratory of Nuclear Physics and Technology, Peking University, Beijing 100871, China}
\affiliation[b]{Department of Engineering Physics, Tsinghua University, Beijing 100084, China}

\emailAdd{yuanzy@pku.edu.cn}

\abstract{%
Precise measurement of the absolute light yield (LY) of scintillators has long been limited by systematic effects inherent in realistic readout geometries. Large-angle incidence, multiple reflections inside the optical housing, and refractive-index mismatch at the coupling interface all introduce biases that cannot be removed by a simple conversion based on the detector's nominal quantum efficiency. To address this problem, we present a correction method that combines the Transfer Matrix Method (TMM) with Geant4 optical Monte Carlo simulation. A wave-optics model of the SiPIN surface thin-film stack is used to extract the angle- and wavelength-dependent single-hit detection probability $p_{\mathrm{det}}(\lambda,\theta)$, which is then dynamically coupled into the macroscopic photon transport simulation, achieving a full-chain integration of the microscopic interface optical response with macroscopic geometric light collection. We demonstrate the method using a GAGG:Ce crystal as the test sample. Two types of optical housings --- a high-absorption Absorber and a high-reflection Reflector --- are each combined with air and optical-grease coupling, forming four independent configurations whose overall photon-to-signal conversion efficiencies $\alpha_{\mathrm{SiPIN}}$ span more than a factor of three. Despite the very different optical boundaries, the intrinsic light yields derived from the four configurations show excellent mutual consistency (coefficient of variation $= 1.8\%$). The measured intrinsic light yield of GAGG:Ce is $LY_{\mathrm{int}} = (5.63 \pm 0.10_{\mathrm{spread}} \pm 0.16_{\mathrm{syst}}) \times 10^{4}~\mathrm{ph/MeV}$. The correction framework effectively decouples the systematic influence of complex geometry and interface optics from photon detection, providing a general-purpose scheme for high-precision, traceable scintillator characterization.
}


\arxivnumber{xxxxxx}

\begin{document}
\maketitle
\flushbottom

\section{Introduction}
\label{sec:introduction}

The absolute light yield (LY) of a scintillator is a central parameter characterising its light emission, defined as the number of scintillation photons produced per unit deposited energy, usually expressed in ph/MeV. LY directly determines the photoelectron statistics available to the detector and thus affects key performance metrics such as energy resolution, threshold and detection efficiency. Establishing a traceable, reproducible method for measuring and calibrating absolute light yield that allows meaningful comparison across different experimental conditions is therefore essential for accurate characterisation of scintillator performance.

In practice, the charge or photoelectron count at the detector output does not directly equal the number of photons emitted inside the crystal. From light emission in the crystal to signal readout, several system effects cause the observed value to deviate from the intrinsic LY. These effects are typically coupled and include the following.

First, the geometric light-collection efficiency is highly sensitive to crystal surface finish (polished/ground/roughness), reflector material, and the presence or absence of an air gap (or partial contact) between the crystal and its wrapping. It is often difficult to obtain accurately by analytical means. In typical scintillator readout geometries, the ideal situation is that photons are efficiently confined at the lateral and top surfaces (via total internal reflection or external reflectors) and eventually reach the bottom readout face where they are detected. In real setups, the angular distribution of photons, interface reflection behaviour and reflector absorption losses are coupled in a complex way. For high-refractive-index crystals such as GAGG ($n \approx 1.9$): when an air gap is preserved between the crystal and its wrapping, total internal reflection (TIR) at the crystal--air interface dominates for angles above the critical angle and keeps photons inside the crystal. If the crystal surface is very smooth, TIR preserves the angle of incidence at each reflection, and some photons remain trapped because their angle always lies outside the escape cone of the readout face and cannot be detected. Introducing surface roughness can randomise angles and break these trapped-light modes, allowing more photons to enter the acceptance cone of the readout face---but it also increases the rate at which photons escape the crystal and interact with lossy reflectors, adding absorption losses \cite{kilimchukStudyLightCollection2010, bircherInvestigationCrystalSurface2012, bartonScintillatorGeometrySurface2007}. These competing effects mean that surface roughness can have very different net effects under different boundary conditions, and the optimal surface state cannot be determined by simple rules. The work of Janecek and Moses systematically measured the reflection properties of crystal surfaces and common reflectors (3M Enhanced Specular Reflector (ESR), Teflon, TiO$_2$ coatings, etc.), showing that real reflection behaviour differs significantly from simplified assumptions such as ``ideal specular'' or ``ideal diffuse'' reflection \cite{janecekMeasuringLightReflectance2008, janecekSimulatingScintillatorLight2010a, janecekOpticalReflectanceMeasurements2008}; Janecek also provided reflectance spectra and thickness dependence for these reflectors in the 250--800~nm range \cite{janecekReflectivitySpectra2012}. This complexity makes Monte Carlo optical simulation necessary for predicting light-collection efficiency under such geometries, and also implies that simulation results are highly sensitive to the choice of surface model and reflector parameters.

Second, there is a fundamental mismatch between detector quantum efficiency and real measurement conditions. The quantum efficiency (QE) of a photodetector is defined as the ratio of primary charge carriers produced (photoelectrons for a PMT, or electron--hole pairs for a silicon detector) to the number of incident photons. Experimentally, QE is usually obtained from spectral responsivity measurements: using collimated light from a monochromator, the response of a reference detector calibrated by a metrology institute and that of the device under test are measured alternately in the same optical path (the ``substitution method''), yielding the quantum efficiency curve of the device at each wavelength \cite{larasonSpectroradiometricDetectorMeasurements}. This standard measurement condition---collimated light, near-normal incidence, air medium---differs significantly from the actual readout environment of a scintillator. de~Haas and Dorenbos analysed this issue systematically \cite{dehaasMeasuringAbsoluteLight2005a, dehaasAdvancesYieldCalibration2008}, pointing out at least three sources of bias in scintillator measurements: (1)~scintillation photons originate inside the crystal and reach the detector after multiple reflections, so the angular distribution of incidence is much broader than the near-normal condition in standard QE measurements, and the angular dependence of QE (mainly through interference in the entrance-layer coatings) is non-negligible at large angles; (2)~Fresnel reflection at the detector surface returns some photons into the crystal and reflector cavity, where they may hit the detector again after further reflection---this multiple-hit effect changes the final detection probability per photon and is often neglected in simplified treatments; (3)~when the coupling medium changes from air to optical grease, the change in refractive index on the incident side significantly affects the effective transmittance of the detector's anti-reflection coating; together with the broad angular distribution, this means the standard QE curve is no longer directly applicable to the real geometry. Direct use of manufacturer QE for correction introduces systematics that are difficult to quantify.

Third, crystal self-absorption is path-length dependent, while encapsulation details (e.g.\ coupling layer thickness, contact fraction, presence of air gaps) are often difficult to control and reproduce precisely, making direct comparison between different samples or batches problematic.

To address these challenges, this paper takes the Hamamatsu S3590 series silicon PIN photodiodes (SiPIN) as an example and establishes a traceable ``charge $\rightarrow$ photon number'' calibration method under realistic measurement geometry. Specifically, the transfer-matrix method (TMM) is used to model the optical thin films on the SiPIN surface, converting the manufacturer's standard QE into an angle- and wavelength-dependent ``single-hit detection probability'' $p_{\text{det}}(\lambda,\theta)$; this is then integrated into a Geant4 optical Monte Carlo simulation to track the full chain of wide-angle incidence, multiple hits, and photon transport in the crystal and reflector, yielding the overall photon-to-signal conversion efficiency $\alpha_{\text{SiPIN}}$ under the real geometry (i.e.\ the total probability that a scintillation photon ultimately produces collectable charge carriers in the SiPIN, as defined in equation~\eqref{eq:alpha_sipin_def}). On the experimental side, the S3590-08 (with Epoxy window) is used for scintillation light readout, while the structurally simpler S3590-09 (without window) serves as the calibration reference for the thin-film optical model; the response of the S3590-08 under real incidence conditions is obtained by extrapolation (see section~\ref{sec:tmm}).

To validate the method, two optically distinct enclosure configurations are designed for the crystal wrapping. The Absorber has low-reflectivity absorbing coating on its inner walls; most photons not directed straight at the readout face are absorbed at the walls, so detection efficiency is low but insensitive to encapsulation details. The Reflector has high-reflectivity TiO$_2$ coating on its inner walls; photons have a high probability of reaching the readout face after multiple reflections in the cavity, so detection efficiency is much higher but depends on the precise value of the wall reflectivity $\rho$. For each enclosure type, two coupling modes are used between the crystal exit face and the SiPIN entrance face: Air (air gap) and Grease (optical grease, $n \approx 1.46$). Air coupling is limited by TIR at the crystal--air interface, so only photons at small angles can exit; grease coupling allows a wider exit cone because the refractive index is closer to that of the crystal. The $\alpha_{\text{SiPIN}}$ values for these $2\times 2 = 4$ configurations span roughly a factor of three in dynamic range.

The wall reflectivity of the Absorber configuration is given by the supplier's process specification and does not rely on any scintillator measurement data; it can therefore serve as an independent calibration reference. The key unknown for the Reflector configuration is the effective reflectivity $\rho_{\text{eff}}$: since the experimental setup does not allow in-situ reflectivity measurement and coating performance is highly sensitive to process and cavity geometry, the ratio $\mathcal{R}_{\mathrm{A-G}}$ of signal peak positions when switching between Air and Grease coupling in the same Reflector geometry is used to constrain this parameter. The physical basis is that the coupling medium changes the acceptance range of exit angles, thereby changing the contribution of multiply reflected photons to the signal, so that $\mathcal{R}_{\mathrm{A-G}}$ and $\rho$ are monotonically related and $\rho_{\text{eff}}$ can be uniquely determined from the measured value. If the Absorber and Reflector paths, which use completely independent parameter sources, yield consistent intrinsic light yield $LY_{\text{int}}$, this indicates that the calibration model is reliable across different optical boundary conditions.

This paper studies a $5\times5\times5$~mm$^3$ GAGG:Ce (Gd$_3$Al$_2$Ga$_3$O$_{12}$:Ce) cubic sample. Its emission spectrum matches the high-sensitivity band of silicon detectors well, and material parameters (emission spectrum, transmittance) can be obtained from experiment. The Geant4 simulation uses the UNIFIED model parameter $\sigma_{\text{surf}}$ to characterise surface roughness; a sensitivity analysis is presented in section~\ref{sec:discussion}. The paper is organised as follows: section~\ref{sec:simulation} describes the optical simulation framework based on TMM and Geant4; section~\ref{sec:experiment} describes the experimental design and electronics calibration; section~\ref{sec:results} derives the intrinsic light yield using the dual-path strategy and presents cross-configuration consistency; section~\ref{sec:discussion} analyses calibration precision via sensitivity scans and systematic uncertainty evaluation, and discusses the influence of encapsulation boundary conditions; section~\ref{sec:conclusions} gives the conclusions.

\section{Simulation}
\label{sec:simulation}

This section describes the optical simulation approach used for absolute light yield calibration. SiPIN detector surfaces are typically covered by multilayer thin-film structures on the micron or sub-micron scale (e.g., SiO$_2$/Si$_3$N$_4$ anti-reflection (AR) coatings), whose thicknesses are comparable to the incident wavelength and give rise to strong optical interference. This interference makes the detector QE depend not only on wavelength but also on incident angle and polarisation. Geant4 optical simulation, however, is based on geometrical optics and cannot directly model such wave-optical effects. To address this, a two-step modelling strategy is adopted: first, a wave-optical model of the SiPIN surface is built using TMM to compute its transmission and absorption as a function of wavelength and incident angle; these results are then stored in a look-up table and used during Geant4 photon transport, coupling macroscopic ray tracing with microscopic wave-optical behaviour. This section first presents the TMM modelling and validation of the SiPIN surface optical response (section~\ref{sec:tmm}), then the implementation of the Geant4 optical transport framework (section~\ref{sec:geant4_optical}).

\subsection{SiPIN surface optical response modelling}
\label{sec:tmm}

Vendor QE curves are typically quoted under standard test conditions (air incidence, near-normal incidence, collimated beam) \cite{dehaasAdvancesYieldCalibration2008}\footnote{Hamamatsu S3590 series product page: \url{https://hep.hamamatsu.com/eu/en/products/S3590-09.html}.}. In the scintillator readout geometry considered here, photons emerge from a high-refractive-index crystal (GAGG:Ce, $n \approx 1.9$), pass through air or optical grease ($n \approx 1.46$), and reach the SiPIN surface with a broad angular distribution; they may also hit the detector multiple times after reflection. The vendor QE curve therefore cannot be used directly as the ``single-hit detection probability'' in the simulation.

To address this, the SiPIN surface is modelled with thin-film optics: once the layer stack parameters are calibrated against the vendor QE curve, TMM can compute the detection probability $p_{\text{det}}(\lambda, \theta)$ for any incident medium of known refractive index, as a function of wavelength and incident angle, and the model is validated by comparison with the vendor QE curve. The Hamamatsu S3590 SiPIN series reaches 95\%--97\% QE in the 600--700~nm range, well above the theoretical limit for bare silicon in air (about 65\%--70\%), consistent with an efficient AR coating on the device surface.

\begin{figure}[htbp]
    \centering
    \includegraphics[width=0.95\textwidth]{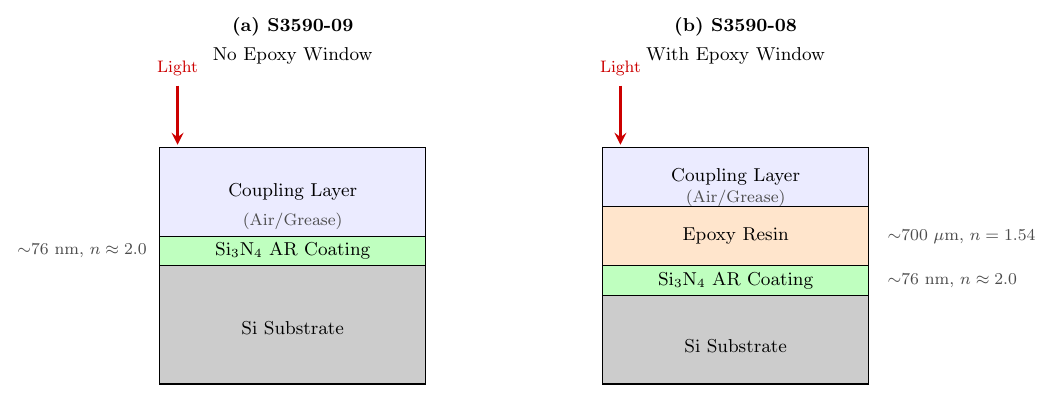}
    \caption{Schematic of the SiPIN surface layer stack. (a) S3590-09: incident medium $\rightarrow$ Si$_3$N$_4$ AR coating $\rightarrow$ Si substrate; (b) S3590-08: incident medium $\rightarrow$ Epoxy window ($\sim$700~$\mu$m) $\rightarrow$ Si$_3$N$_4$ AR coating $\rightarrow$ Si substrate.}
    \label{fig:layer_structure}
\end{figure}

For a single-layer quarter-wave AR coating, thin-film optics gives the zero-reflection condition as $n_{\text{AR,opt}} = \sqrt{n_0 \cdot n_{\text{sub}}}$, i.e.\ the film refractive index equals the geometric mean of the incident and substrate indices \cite{macleodThinFilmOpticalFilters2010}. For air incidence ($n_0 = 1$) onto a silicon substrate ($n_{\text{Si}} \approx 3.5$--$4.0$), the optimal film index is therefore about $1.9$--$2.0$. Accordingly, single-layer Si$_3$N$_4$ with $n \approx 2.0$--$2.1$ is used as an equivalent model for the AR coating; once the layer stack parameters are fixed, TMM correctly predicts the response for any incident medium. The resulting layer stack structure is shown in figure~\ref{fig:layer_structure}: the S3590-09 (without resin window) is simplified to ``incident medium $\rightarrow$ Si$_3$N$_4$ ($d_{\text{AR}}$) $\rightarrow$ Si substrate''; the S3590-08, following the vendor datasheet, adds a 700~$\mu$m epoxy protective layer above the AR coating.
    
TMM is used to compute the optical response of the multilayer stack \cite{macleodThinFilmOpticalFilters2010, byrnesMultilayerOpticalCalculations2020}. The transfer matrix $M$ is obtained by multiplying the interface and propagation matrices for each layer; the complex reflection coefficient $r = M_{21}/M_{11}$ and power reflectance $R = |r|^2$ are then derived, with the average of s- and p-polarisation taken for unpolarised light. Silicon (Si) has significant absorption in the visible; the model therefore uses a complex refractive index $\tilde{n} = n + ik$, with the branch of the complex refraction angle in Snell's law chosen so that the electromagnetic wave decays in the absorbing medium. Optical constants are taken from standard datasets: Si from Green (2008), Si$_3$N$_4$ from Philipp (1973) \cite{greenSelfconsistentOpticalParameters2008, philippOpticalPropertiesSilicon1973}. The detection probability is defined as
\begin{equation}
    p_{\text{det}}(\lambda, \theta) = \eta_{\text{IQE}} \cdot [1 - R(\lambda, \theta)],
    \label{eq:pdet_def}
\end{equation}
where $\theta$ is the angle of incidence measured from the surface normal, $\eta_{\text{IQE}}$ is the internal quantum efficiency (accounting for carrier collection, dead layer, and other internal losses) and $R(\lambda, \theta)$ is the entrance reflectance from TMM.

The S3590-09 has no epoxy window and a simpler surface structure, making it suitable as a calibration reference. For a single-layer film, the TMM entrance reflectance $R(\lambda,\theta;d_{\text{AR}})$ depends explicitly on the film thickness $d_{\text{AR}}$. At normal incidence:
\begin{equation}
    R(\lambda,0;d_{\text{AR}}) = \left|\frac{r_{01}+r_{12}\,e^{2i\delta}}{1+r_{01}r_{12}\,e^{2i\delta}}\right|^2,\quad
    \delta = \frac{2\pi n_{\text{AR}} d_{\text{AR}}}{\lambda},
    \label{eq:tmm_single_layer}
\end{equation}
where $r_{01}$ and $r_{12}$ are the Fresnel amplitude reflection coefficients at the incident--AR and AR--Si interfaces, and $\delta$ is the single-pass phase in the film. The datasheet QE curve $QE_{09}^{\mathrm{(vendor)}}(\lambda)$ is used as a constraint and equated to the single-incidence detection probability at $\theta=0$:
\begin{equation}
    QE_{09}^{\mathrm{(model)}}(\lambda; d_{\text{AR}}, \eta_{\mathrm{IQE}})
    \equiv p_{\mathrm{det}}(\lambda, 0)
    = \eta_{\mathrm{IQE}}\,[1-R_{09}^{\mathrm{(TMM)}}(\lambda,0;d_{\text{AR}})].
\end{equation}
Denoting the datasheet sampling wavelengths by $\Lambda=\{\lambda_j\}$ and defining $A_j(d_{\text{AR}})\equiv 1-R_{09}^{\mathrm{(TMM)}}(\lambda_j,0;d_{\text{AR}})$, the fit minimises
\begin{equation}
    \min_{d_{\text{AR}},\eta_{\mathrm{IQE}}}\ \sum_{j}\left(QE_{09}^{\mathrm{(vendor)}}(\lambda_j)-\eta_{\mathrm{IQE}}A_j(d_{\text{AR}})\right)^2.
\end{equation}

The S3590-08 has an epoxy protective window of about 700~$\mu$m thickness (refractive index $n_{\text{epoxy}} \approx 1.54$). Because this layer is much thicker than the optical wavelength, it is treated as an incoherent thick layer. Optical losses in thick polymer layers arise mainly from Rayleigh scattering ($\propto \lambda^{-4}$) and from absorption and scattering by impurities and defects\footnote{Epoxy Technology, Inc., ``Tech Tip 18: Understanding Optical Properties of Epoxy Applications,'' \url{https://www.epotek.com/docs/en/Related/Tech\%20Tip\%2018\%20Understanding\%20Optical\%20Properties\%20of\%20Epoxy\%20Applications.pdf}.}. Since the fit is performed over a narrow band (450--700~nm), the window transmittance $T_{\text{epoxy}}(\lambda)$ is approximated as $T_{\text{epoxy}}(\lambda) \approx a + b \cdot (\lambda - \lambda_0)$ with $\lambda_0 = 550$~nm, and this factor is applied on top of the same entrance reflectance calculation as for the S3590-09:
\begin{equation}
    QE_{08}^{\mathrm{(model)}}(\lambda; d_{\text{AR}}, \eta_{\mathrm{IQE}}, a, b)
    = T_{\mathrm{epoxy}}(\lambda)\cdot \eta_{\mathrm{IQE}}\,[1-R_{08}^{\mathrm{(TMM)}}(\lambda,0;d_{\text{AR}})].
\end{equation}
A two-step joint fit is used: first $(d_{\text{AR}},\eta_{\mathrm{IQE}})$ is fitted using only S3590-09 QE data; these are then fixed and $(a,b)$ is fitted using S3590-08 QE data.

\begin{figure}[htbp]
    \centering
    \begin{minipage}[b]{0.48\textwidth}
        \centering
        \includegraphics[width=\textwidth]{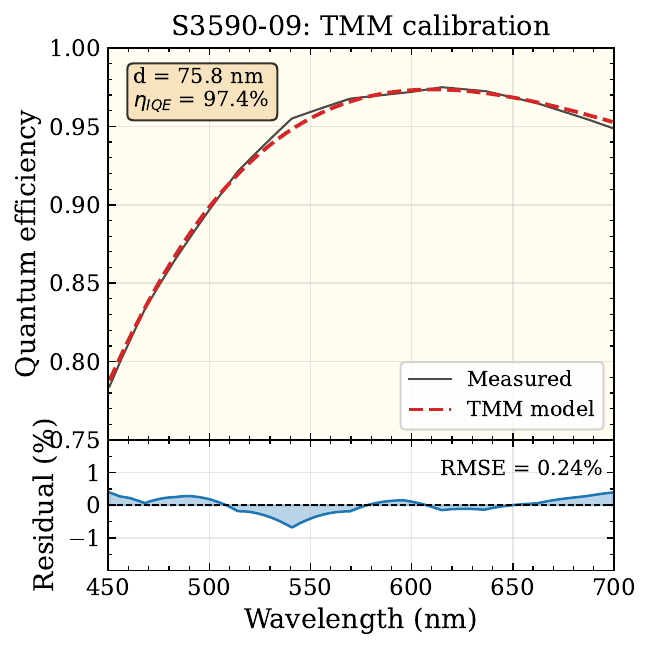}
        \subcaption{S3590-09 (without window)}
        \label{fig:qe_calibration_09}
    \end{minipage}
    \hfill
    \begin{minipage}[b]{0.48\textwidth}
        \centering
        \includegraphics[width=\textwidth]{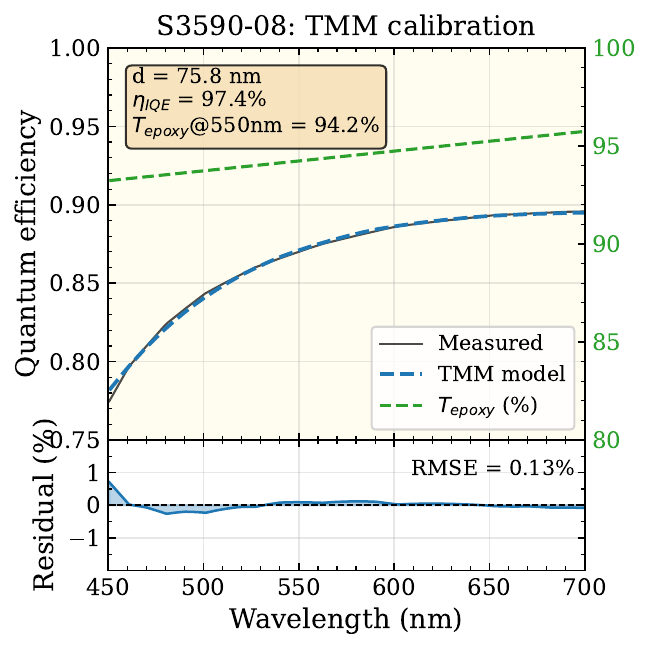}
        \subcaption{S3590-08 (with Epoxy window)}
        \label{fig:qe_calibration_08}
    \end{minipage}
    \caption{SiPIN TMM model calibration results. Each subfigure shows the vendor QE curve vs.\ TMM prediction (top) and residuals (model $-$ measured) (bottom). (a)~S3590-09: fitted parameters $d_{\text{AR}} = 75.75$~nm, $\eta_{\text{IQE}} = 97.4\%$. (b)~S3590-08: shared $(d_{\text{AR}}, \eta_{\text{IQE}})$; green dashed line is the fitted Epoxy transmittance $T_{\text{epoxy}}(\lambda)$, about 94.2\% at 550~nm.}
    \label{fig:qe_calibration}
\end{figure}

Figure~\ref{fig:qe_calibration} shows the fit results. For the S3590-09, the model agrees with the vendor QE data over the fit range (root-mean-square error (RMSE) about 0.24\%); the fitted internal quantum efficiency $\eta_{\text{IQE}}\approx97.4\%$ is consistent with values reported for silicon photodiodes \cite{gentileInternalQuantumEfficiency2010a}. For the S3590-08, the fit gives $b > 0$ (lower transmittance at short wavelengths), consistent with stronger Rayleigh scattering losses at shorter wavelengths; the right-hand axis in figure~\ref{fig:qe_calibration}(b) shows $T_{\text{epoxy}} \approx 94.2\%$ at 550~nm, corresponding to about 6\% absorption/scattering loss. The adopted calibration parameters are $d_{\text{AR}} = 75.75$~nm and $\eta_{\text{IQE}} = 97.39\%$.

\begin{figure}[htbp]
    \centering
    \includegraphics[width=\textwidth]{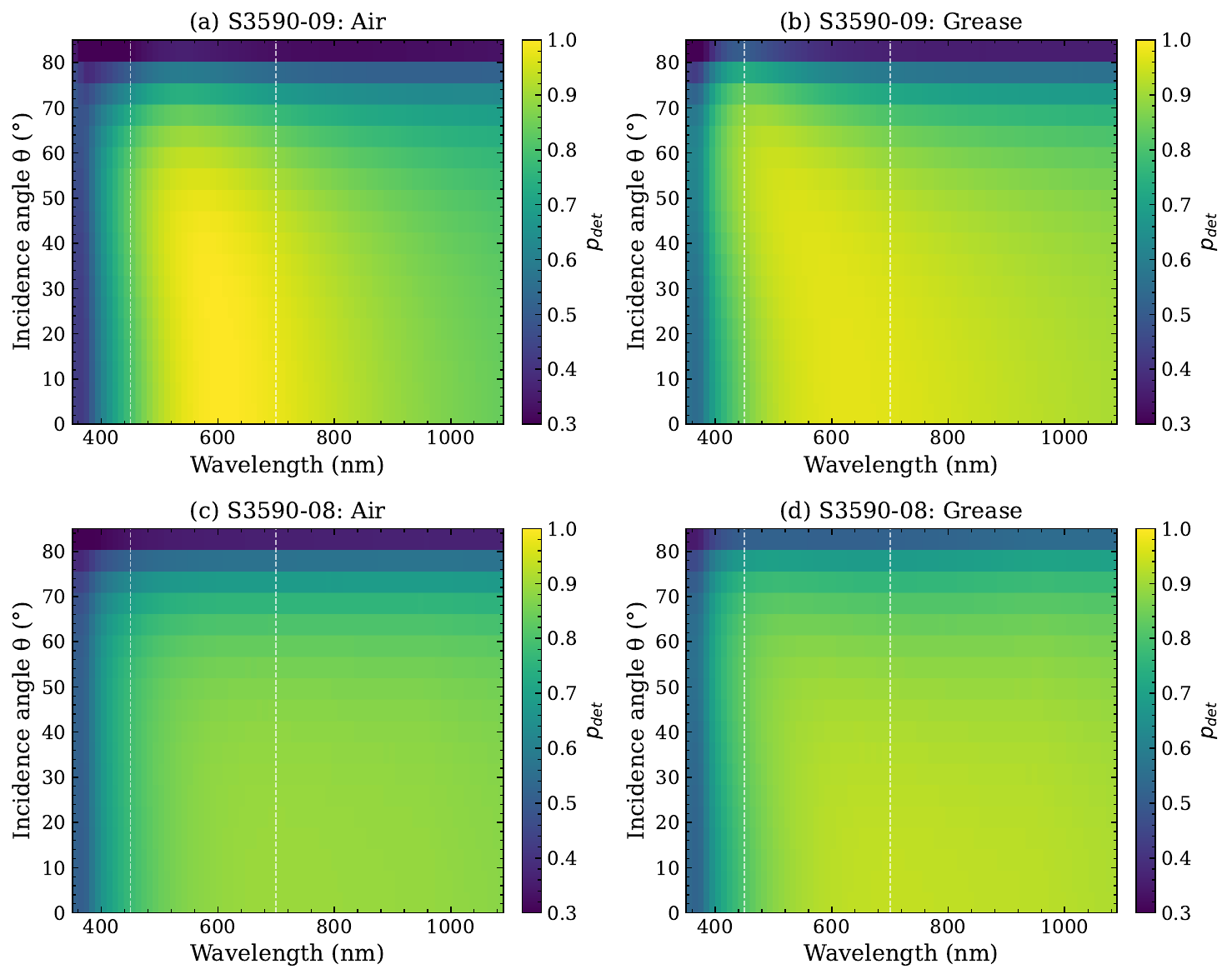}
    \caption{Detection probability $p_{\text{det}}(\lambda, \theta)$ heat maps. (a) S3590-09, air incidence; (b) S3590-09, grease incidence; (c) S3590-08, air incidence; (d) S3590-08, grease incidence. White dashed lines mark the GAGG emission band (450 and 700~nm).}
    \label{fig:pdet_heatmap_all}
\end{figure}

With the calibrated layer stack parameters fixed, TMM can compute the detection probability for any incident medium of known refractive index, as a function of wavelength and incident angle. Figure~\ref{fig:pdet_heatmap_all} shows two-dimensional heat maps of $p_{\text{det}}(\lambda, \theta)$ for the S3590-09 and S3590-08 under air and grease incidence. Near normal incidence ($\theta < 30^\circ$), the S3590-09 performs slightly better in air than in grease, since the AR coating is optimised for the air--Si interface; at larger angles ($\theta > 50^\circ$), grease gives higher detection probability because its refractive index is closer to Si$_3$N$_4$, increasing the critical angle and slowing the rise in reflectance. The S3590-08 has slightly lower overall detection probability than the S3590-09 due to extra absorption in the epoxy window; in the GAGG emission band (450--700~nm), the detection probability remains high ($> 0.8$). The $p_{\text{det}}(\lambda, \theta)$ tables are generated on a grid with $\Delta\lambda = 10$~nm and $\Delta\theta = 5^\circ$, stored in CSV format, and loaded at Geant4 initialisation for bilinear interpolation at boundary interactions.

The TMM equivalent model uses a simplified layer stack (single-layer Si$_3$N$_4$ plus linear $T_{\text{epoxy}}$). Its model uncertainty is estimated from three sources. The statistical uncertainty of the fitted parameters $(d_{\text{AR}},\eta_{\text{IQE}},a,b)$ propagates to the $p_{\text{det}}(\lambda,\theta)$ grid and contributes about $0.05\%$, indicating high statistical stability. The effect of the fit wavelength range is assessed by varying the range boundaries and comparing $p_{\text{det}}$ results when extrapolating to arbitrary angles and media; this contributes about $0.24\%$. In addition, the single-layer AR assumption introduces a structural bias relative to the real multilayer stack that cannot be fully removed even with optimal parameters; the joint fit residual $\text{RMSE}_{\text{joint}}/\overline{QE}_{\text{band}} \approx 0.27\%$ is used as a conservative estimate of this structural error. Combining these three contributions in quadrature gives an overall model uncertainty of about $0.37\%$.

\subsection{Geant4 optical simulation}
\label{sec:geant4_optical}

\begin{figure}[htbp]
    \centering
    \includegraphics[width=0.65\textwidth]{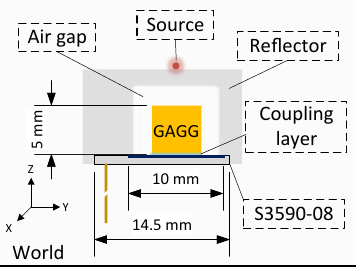}
    \caption{Schematic of the Geant4 simulation geometry. The crystal is surrounded by a reflector or absorber housing; the bottom face is coupled to the SiPIN via a coupling layer.}
    \label{fig:sim_geometry}
\end{figure}

Geant4 v11.2.2 is used for optical photon transport \cite{agostinelliGeant4aSimulationToolkit2003a, allisonRecentDevelopmentsGeant42016a}. The reference geometry matches the experimental reflector housing (figure~\ref{fig:sim_geometry}): a $5 \times 5 \times 5$~mm$^3$ GAGG:Ce cubic crystal is placed inside the housing, with an air gap of about 2\,mm between the crystal sides and top and the housing walls; the crystal bottom face is coupled to the SiPIN detector via a coupling layer (air or grease). The housing bottom inner diameter matches the SiPIN outline for mechanical alignment and reproducible positioning across measurements. The source is placed 5\,mm above the housing, with emission directions sampled uniformly over the sphere; each run ensures more than 30\,000 events with energy deposited in the crystal. The optical physics processes enabled include bulk absorption, Rayleigh scattering, boundary reflection/refraction, and scintillation.

\begin{figure}[htbp]
    \centering
    \includegraphics[width=0.75\textwidth]{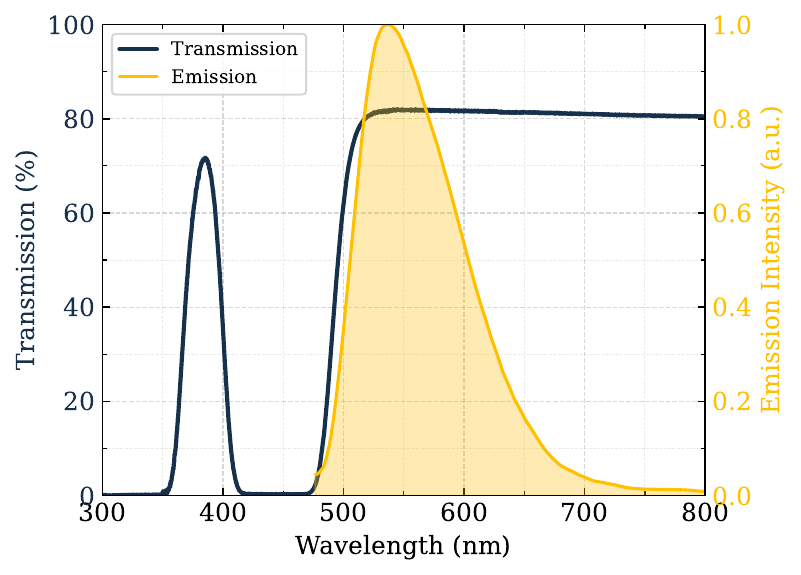}
    \caption{Transmittance and emission spectrum of the GAGG:Ce sample ($5\times5\times5$~mm$^3$).}
    \label{fig:gagg_optical}
\end{figure}

Material optical parameters are set as follows. The GAGG:Ce refractive index uses the $n(\lambda)$ curve from Kozlova et al.\ \cite{kozlovaOpticalCharacteristicsSingleCrystal2018} with interpolation; the emission spectrum is measured with an Edinburgh FLS1000 fluorimeter, with peak wavelength about 530~nm; the transmittance is measured with an Edinburgh DS5 UV--visible spectrophotometer (200--800~nm), and the absorption length $L_{\text{abs}}(\lambda)$ is derived after accounting for double-surface reflection at the air--crystal interface. Figure~\ref{fig:gagg_optical} shows the measured transmittance and emission spectrum and it drops sharply for $\lambda < 450$~nm, corresponding to the Ce$^{3+}$ 4f--5d absorption; the emission peak is at $\sim$530~nm, with the main band between 450 and 700~nm, matching the SiPIN high-sensitivity region. The housing walls use a Lambertian diffuse reflection model with reflectivity $\rho$ as a tunable parameter. Two housings with the same geometry but different wall treatments are used in the experiment: Reflector and Absorber (section~\ref{sec:sample_config}). The coupling layer refractive index is set to $n = 1.0$ for air or $n = 1.46$ for grease\footnote{EJ-550 optical grease product page and datasheet: \url{https://eljentechnology.com/products/accessories/ej-550-ej-552}.}, with transmittance from the datasheet (about 98.8\% for 0.1~mm thickness at 450--500~nm) used as reference. The SiPIN angle--wavelength dependent response is taken from the $p_{\text{det}}(\lambda,\theta)$ look-up table computed in section~\ref{sec:tmm}; the ceramic substrate, epoxy window, and silicon sensitive layer are modelled according to the datasheet, while pins and shielding are not included.

Crystal surface roughness is described by the $\sigma_{\text{surf}}$ parameter of the Geant4 UNIFIED model (implemented as \texttt{sigma\_alpha}, in rad). This parameter sets the angular spread of the micro-surface normal relative to the macro-normal; $\sigma_{\text{surf}} \to 0$ corresponds to a specular surface, and larger $\sigma_{\text{surf}}$ increases angular randomisation \cite{gumplingerOpticalPhotonProcesses}. $\sigma_{\text{surf}}$ is a parameter in the geometrical-optics surface model and is not equivalent to machining roughness metrics such as Ra/Rq. In Geant4 applications, Jungmann et al.\ use $\sigma_{\text{surf}}\approx 0.02$\,rad for polished surfaces and $\sigma_{\text{surf}}\approx 0.2$\,rad for unpolished surfaces, and discuss the range $\sigma_{\text{surf}}\in[0.01,0.3]$\,rad for uncertainty estimation \cite{jungmannDaJungmann2011,janecekSimulatingScintillatorLight2010a}. Here, $\sigma_{\text{surf}} = 0.02$~rad is used as the nominal value for polished surfaces ($\sigma_{\text{surf}} = 0.2$~rad for ground), and a sensitivity scan over 0.01--0.3~rad is performed to cover the range from near-specular to clearly rough.

When an optical photon reaches the ``coupling medium $\rightarrow$ SiPIN'' boundary, the simulation computes its wavelength $\lambda$ and incident angle $\theta$, looks up the single-hit detection probability $p_{\text{det}}(\lambda, \theta)$ from the TMM pre-computed table, and performs a random draw: if the photon is ``detected'', it is counted and the track is terminated; otherwise it is reflected according to the reflection model and continues propagating. This online decision naturally includes the angular distribution and multiple-hit enhancement in a single run.

In this framework, a ``detected'' photon is one that successfully generates collectable primary carriers (electron--hole pairs) in the SiPIN entrance structure and thus contributes to the measured charge signal. The simulation records the number of detected photons $N_{\text{det}}$ per event and defines the equivalent electron--hole pair yield as $N_{ehp}^{\text{(sim)}}\equiv N_{\text{det}}$, so that observables consistent with the experimental peak ratio can be constructed (section~\ref{sec:results}).

For a scintillation photon to be detected by the SiPIN, it must pass through all of the following steps without loss: (1)~propagate in the crystal without self-absorption (described by the wavelength-dependent absorption length $L_{\text{abs}}(\lambda)$); (2)~at crystal surfaces, refraction or reflection with micro-surface roughness must not deflect it outside the solid angle reaching the readout face (parameterised by $\sigma_{\text{surf}}$); (3)~exit through the crystal side and top faces at the crystal--air interface, directly or after multiple reflections; (4)~when hitting the housing walls, not be absorbed (described by reflectivity $\rho$ and the Lambertian diffuse reflection model); (5)~pass through the coupling layer (air or grease) to the SiPIN entrance, traverse the AR coating stack at wavelength $\lambda$ and incident angle $\theta$, and generate carriers in silicon (described by the TMM $p_{\text{det}}(\lambda,\theta)$ in section~\ref{sec:tmm}). These steps are tracked explicitly photon-by-photon in the simulation; photons may undergo multiple reflections inside the crystal and hit the SiPIN surface several times, with each hit judged independently using the current $(\lambda,\theta)$.

Combining these effects, the overall photon-to-signal conversion efficiency is defined as
\begin{equation}
    \alpha_{\text{SiPIN}} \equiv \frac{N_{\text{det}}}{N_{\gamma}} ,
    \label{eq:alpha_sipin_def}
\end{equation}
where $N_{\gamma}$ is the initial number of scintillation photons and $N_{\text{det}}$ is the number ultimately classified as detected. $\alpha_{\text{SiPIN}}$ is the final output of the optical simulation and links the intrinsic light yield $LY_{\text{int}}$ (an inherent property of the scintillator) to the experimentally measurable electron--hole pair yield $Y_{ehp}$: $LY_{\text{int}} = Y_{ehp} / \alpha_{\text{SiPIN}}$ (derivation in section~\ref{sec:results}).

In the implementation, the custom boundary logic uses a ``kill and respawn'' strategy\footnote{Modifying the photon momentum directly in \texttt{UserSteppingAction} can be overwritten by \texttt{G4OpBoundaryProcess} later in the same step. The current track is therefore marked \texttt{fStopAndKill}, and a new photon with the updated momentum direction is added to the secondary queue, avoiding conflict with the built-in boundary process.}, so that the reflection decision does not conflict with the Geant4 boundary process.

Within this simulation framework, most inputs to $\alpha_{\text{SiPIN}}$ are fixed from independent sources: the TMM layer stack parameters $(d_{\text{AR}}, \eta_{\text{IQE}})$ come from the vendor QE fit (section~\ref{sec:tmm}); the GAGG refractive index $n(\lambda)$, absorption length $L_{\text{abs}}(\lambda)$, and emission spectrum from separate spectroscopic measurements; geometry from the actual setup; and surface roughness $\sigma_{\text{surf}}$ from the literature value for polished surfaces. For the Absorber configuration, absorption dominates at the low reflectivity of the coating, and $\alpha_{\text{SiPIN}}$ is insensitive to the exact value (section~\ref{sec:uncertainties}); the coating vendor reflectivity $\rho \approx 0.6\%$ is used directly in the simulation.

The only key parameter that cannot be fixed independently is the wall reflectivity $\rho$ for the Reflector configuration. The performance of high-reflectivity coatings depends on coating process and cavity geometry, and in a small housing it is difficult to reach the ideal values reported in the literature; the current setup does not allow in-situ angle-resolved reflectivity measurements. The peak ratio $\mathcal{R}_{\mathrm{A-G}}$ between grease and air coupling for the same Reflector geometry is therefore used to constrain $\rho$ indirectly (section~\ref{sec:results}).

Thus, the Absorber and Reflector configurations use $\rho$ values from entirely different sources and span a dynamic range of about $20\%$--$67\%$, providing a test of the reliability of the calibration method (section~\ref{sec:results}).

\section{Experimental Setup}
\label{sec:experiment}

This section describes the experimental apparatus, data acquisition system, and the measurement procedure for absolute light yield. The experiment is divided into two main steps: first, direct irradiation of the SiPIN with $^{241}$Am establishes the absolute conversion factor from pulse amplitude to electron--hole pair number; subsequently, the photopeak response of a GAGG:Ce crystal under $^{22}$Na excitation is measured in four different encapsulation and coupling configurations. The electronics settings are kept consistent throughout the experiment to ensure that the calibration coefficient can be applied directly to light yield derivation.

\subsection{Sample configuration and hardware system}
\label{sec:sample_config}

The sample under study is a $5 \times 5 \times 5$~mm$^3$ GAGG:Ce scintillation crystal with six optically polished faces. The crystal is placed inside a reflector/absorber housing with an air gap of approximately 2~mm between the crystal and the housing walls; the crystal bottom face is in contact with a Hamamatsu S3590-08 SiPIN photodiode through a coupling layer.

This work employs a $2\times2$ combination of two housing types and two coupling methods, yielding four configurations in total. The Reflector housing is fabricated by white plastic 3D printing with a TiO$_2$ white diffuse reflection coating on the inner walls; the Absorber housing has the same geometry as the Reflector but is fabricated by black plastic 3D printing with a Musou Black Paint ultra-low reflectivity coating applied to the inner walls by air-brush\footnote{Musou Black Paint product page: \url{https://the-black-market.com/products/musou-black-paint}.}, with a manufacturer-specified reflectivity of approximately $0.6\%$. The inner diameter at the bottom of both housings matches the SiPIN outer contour, forming a mechanical stop to ensure housing position consistency across repeated measurements. This $2\times2$ combination yields $\alpha_{\text{SiPIN}}$ spanning approximately $20\%$--$67\%$ of the dynamic range across the four configurations, and forms two independent $\rho$ determination paths---Absorber uses the manufacturer parameter, Reflector is constrained by the experimental ratio (section~\ref{sec:results})---for subsequent cross-validation. Grease coupling uses EJ-550 optical grease ($n \approx 1.46$), with a nominal coupling layer thickness of 50~$\mu$m; dry coupling applies no coupling agent, with the crystal placed directly on the SiPIN Epoxy window.

\begin{figure}[htbp]
\centering
\includegraphics[width=0.85\textwidth]{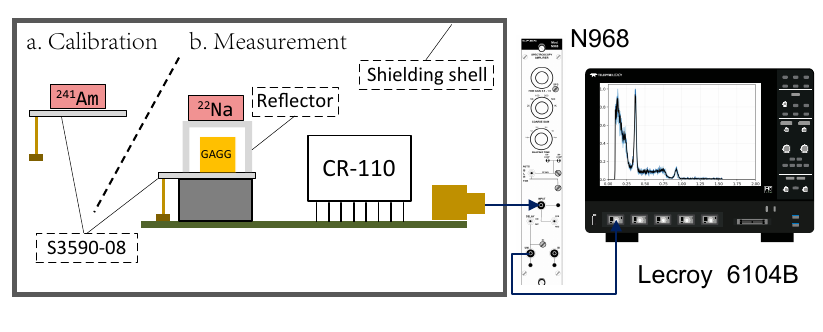}
\caption{Readout chain under two operating conditions. (a) Absolute charge calibration by direct $^{241}$Am irradiation of the SiPIN. (b) Scintillation measurement using a GAGG:Ce crystal under $^{22}$Na excitation. The same electronics chain and settings are used in both cases; the CSA and its input connection are enclosed in a shielding shell to suppress electromagnetic pickup and baseline noise.}
\label{fig:LY_setup}
\end{figure}

The SiPIN operates at 74~V reverse bias. Its output charge pulses are first converted to voltage pulses by a charge-sensitive preamplifier (CSA) Cremat CR-110\footnote{Cremat CR-110 charge-sensitive preamplifier datasheet: \url{https://www.cremat.com/CR-110-R2.2.pdf}.}, then fed to a shaping amplifier CAEN N968\footnote{CAEN N968 shaping amplifier product page: \url{https://www.caen.it/products/n968/}.} for signal shaping and gain adjustment (shaping time set to 2~$\mu$s). The shaped signal is acquired by a Teledyne LeCroy HDO6104B oscilloscope\footnote{Teledyne LeCroy HDO6104B oscilloscope product page: \url{https://www.teledynelecroy.com/oscilloscope/hdo6000b-high-definition-oscilloscopes/hdo6104b}.} in Histogram mode to record the pulse amplitude spectrum. To suppress electromagnetic interference and reduce baseline noise, the front-end CSA and critical circuitry near its input are enclosed in a 5~mm thick, effectively grounded aluminium shielding box. Figure~\ref{fig:LY_setup} shows a schematic of the readout chain.

For scintillation measurements, the $^{22}$Na source is placed above the centre of the housing; the source position is kept fixed when switching coupling configurations. For charge calibration, the scintillation crystal and housing are removed while the circuit connections remain unchanged, and the $^{241}$Am source irradiates the SiPIN directly.

\subsection{Charge calibration and data acquisition}
\label{sec:electrical_calibration}

The absolute charge calibration of the system employs the ``direct irradiation method'': the $^{241}$Am source irradiates the SiPIN directly, and the full-energy peak formed by its $E_{\mathrm{Am}} = 59.54$~keV characteristic photons depositing energy in the Si sensitive layer is used to establish the correspondence between oscilloscope pulse amplitude (V) and equivalent input charge (or electron--hole pair number) \cite{mengDesignPerformanceLargevolume2002a,moszynskiAbsoluteLightOutput1997}. It is assumed that charge collection remains linear within the region of interest (ROI) corresponding to the selected peak position, and that $W_{\mathrm{Si}}$ can be used to convert deposited energy to the mean $N_{ehp}$. The effective Si thickness of the SiPIN used in this experiment is approximately 300~$\mu$m (see datasheet). A Gaussian fit to the full-energy peak yields the peak position $V_{\mathrm{Am}}$, and the linear conversion coefficient from peak position to electron--hole pair number is defined as
\begin{equation}
k_{ehp} \equiv \frac{N_{ehp}}{V-V_0} = \frac{N_{ehp}^{(\mathrm{Am})}}{V_{\mathrm{Am}}-V_0},
\label{eq:keh_cal}
\end{equation}
where $V_0$ is the baseline offset (DC bias level) of the readout chain with no signal input, and $N_{ehp}^{(\mathrm{Am})}$ is the mean number of electron--hole pairs produced by $E_{\mathrm{Am}}$ photons in silicon. The shaping amplifier output baseline in this experiment is stable near zero ($V_0 \approx 0$) and is not written explicitly in subsequent calculations. Under the approximation of complete charge collection, this value is determined by the mean ionisation energy of silicon $W_{\mathrm{Si}}$:
\begin{equation}
N_{ehp}^{(\mathrm{Am})} \simeq \frac{E_{\mathrm{Am}}}{W_{\mathrm{Si}}}, \quad W_{\mathrm{Si}} \approx 3.67~\text{eV/ehp},
\label{eq:neh_am}
\end{equation}
where $W_{\mathrm{Si}}$ is taken from the literature \cite{knollRadiationDetectionMeasurement2010}.

For scintillation measurements, the $^{22}$Na source excites the crystal, and events in the 1274.5~keV photopeak region are selected uniformly across the four configurations used in this work. Denoting the photopeak position as $V_{\mathrm{Na}}$, the corresponding electron--hole pair number is obtained from equation~(\ref{eq:keh_cal}) as
\begin{equation}
N_{ehp}^{(\mathrm{Na})} = k_{ehp}\,V_{\mathrm{Na}},
\label{eq:neh_na}
\end{equation}
and the electron--hole pair yield is defined as
\begin{equation}
Y_{ehp} \equiv \frac{N_{ehp}^{(\mathrm{Na})}}{E_{\mathrm{dep}}^{(\mathrm{Na})}} \quad (\text{ehp/MeV}),
\label{eq:Yeh_from_na}
\end{equation}
where $E_{\mathrm{dep}}^{(\mathrm{Na})}$ is the mean deposited energy in the crystal for events entering the photopeak ROI. For photopeak events, the full energy of the $\gamma$ ray is deposited in the crystal, hence $E_{\mathrm{dep}}^{(\mathrm{Na})} = 1274.5$~keV.

\section{Results}
\label{sec:results}

This section presents the experimental measurement of absolute light yield and the correction derivation. The correction relation from experimentally measured quantities to intrinsic light yield is first recalled. The intrinsic light yield is defined as the number of initial scintillation photons produced per unit deposited energy:
\begin{equation}
    N_{\gamma} = LY_{\text{int}} \cdot E_{\text{dep}} .
    \label{eq:Ngamma_def}
\end{equation}
Section~\ref{sec:experiment} has established the experimentally measurable electron--hole pair yield $Y_{ehp} \equiv N_{ehp}/E_{\text{dep}}$ (equation~\eqref{eq:Yeh_from_na}). Denoting the total probability that an initial scintillation photon ultimately produces a collectable signal in the SiPIN as $\alpha_{\text{SiPIN}}$ (equation~\eqref{eq:alpha_sipin_def}, given by the Geant4+TMM combined simulation), the mean response satisfies
\begin{equation}
    N_{ehp} = N_{\gamma} \cdot \alpha_{\text{SiPIN}} .
    \label{eq:Neh_chain}
\end{equation}
Combining equations~\eqref{eq:Ngamma_def} and~\eqref{eq:Neh_chain}, eliminating $N_{\gamma}$ and dividing by $E_{\text{dep}}$, yields the correction relation:
\begin{equation}
    LY_{\text{int}}
    = \frac{Y_{ehp}}{\alpha_{\text{SiPIN}}} .
    \label{eq:full_correction}
\end{equation}
Here $Y_{ehp}$ is determined entirely by experiment ($^{241}$Am calibration + $^{22}$Na peak position), and $\alpha_{\text{SiPIN}}$ is determined entirely by simulation---the two are obtained independently and combined via the above relation to infer the intrinsic light yield. This section first presents the $^{241}$Am absolute calibration coefficient and the experimental spectra for the four configurations, then uses the Grease/Air peak position ratio in the Reflector configuration to constrain the effective reflectivity $\rho_{\text{eff}}$, and finally combines the four results to derive $LY_{\text{int}}$ and validates the method reliability through cross-configuration consistency checks.

\subsection{$^{241}$Am charge calibration and experimental spectra}

Absolute charge calibration is performed by direct irradiation of the SiPIN with an $^{241}$Am source. The number of electron--hole pairs produced in silicon by the 59.54~keV $\gamma$ rays from $^{241}$Am is determined by the mean ionisation energy: $N_{ehp}^{(\mathrm{Am})} = E_{\mathrm{Am}}/W_{\mathrm{Si}} = 16\,223$ (equation~\eqref{eq:neh_am}). A Gaussian fit to the full-energy peak yields the measured peak position $V_{\mathrm{Am}} = 0.3481 \pm 0.0007$~V, giving the absolute charge calibration coefficient:
\begin{equation}
    k_{ehp} = \frac{N_{ehp}^{(\mathrm{Am})}}{V_{\mathrm{Am}}} = 46\,605~\text{ehp/V} .
\end{equation}
This coefficient is used to convert the peak voltage from subsequent scintillation measurements into the number of detected electron--hole pairs.

\begin{figure}[htbp]
    \centering
    \includegraphics[width=\textwidth]{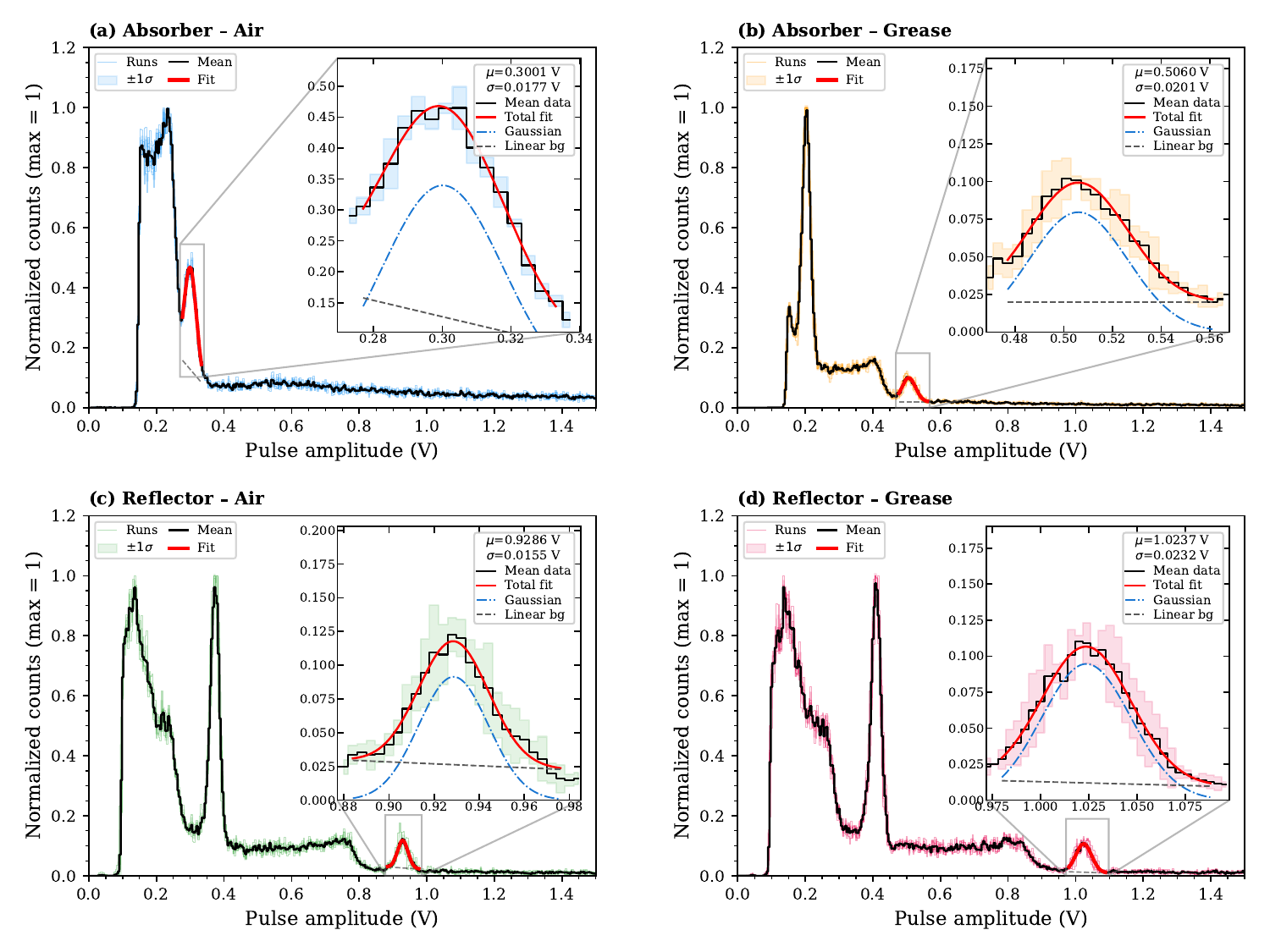}
    \caption{Na-22 pulse amplitude spectra for the four measurement configurations: (a) Absorber--Air, (b) Absorber--Grease, (c) Reflector--Air, (d) Reflector--Grease; all panels use the 1274.5~keV photopeak ($N_{\text{rep}}=5$ for each configuration). Light-coloured lines show individual max-normalised runs; bold black line is the mean; shaded band is $\pm 1\sigma$. Red curves are the Gaussian + linear fit to the mean spectrum.}
    \label{fig:exp_spectra_fits}
\end{figure}

The $5\times5\times5$~mm$^3$ GAGG:Ce sample is then placed in the Absorber ($\rho=0.6\%$) and Reflector housings, each with Air and Grease coupling conditions, and scintillation light is measured with a $^{22}$Na source ($N_{\text{rep}}=5$ repeats per configuration). Figure~\ref{fig:exp_spectra_fits} shows the 1274.5~keV photopeak pulse amplitude spectra for the four configurations. The peak positions span from $0.30$~V (Absorber--Air) to $1.02$~V (Reflector--Grease), a range exceeding a factor of three, directly reflecting the significant effect of encapsulation and coupling conditions on light collection efficiency. All subsequent analysis is based on the Gaussian-fit mean of the 1274.5~keV photopeak for each configuration.

\subsection{Determination of effective reflectivity}
\label{sec:rho_constraint}

\begin{figure}[htbp]
    \centering
    \includegraphics[width=\textwidth]{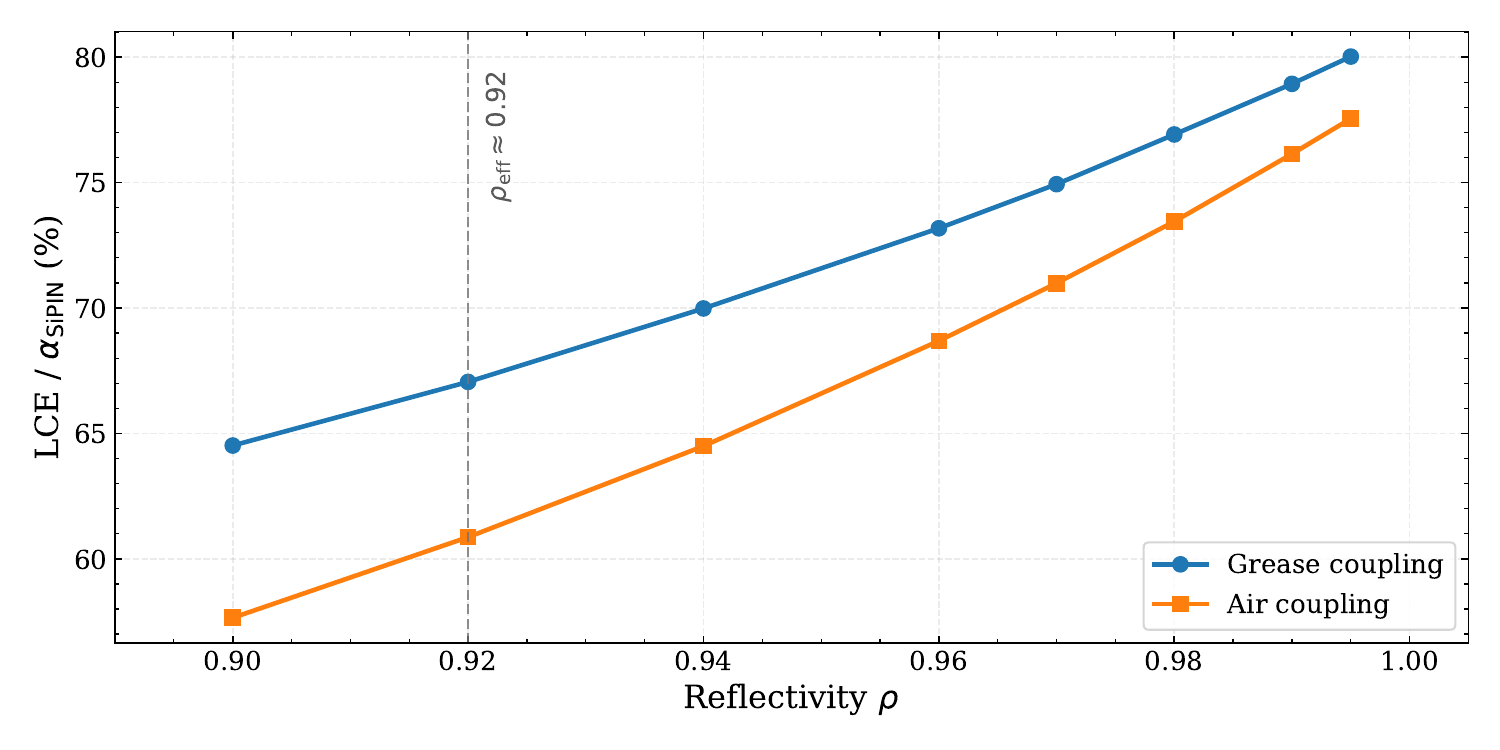}
    \caption{$\alpha_{\text{SiPIN}}$ as a function of reflector wall reflectivity $\rho$. Both Grease and Air coupling conditions are shown; the dashed line marks the effective reflectivity $\rho_{\text{eff}} \approx 0.92$ constrained by the experimental $\mathcal{R}_{\mathrm{A-G}}$.}
    \label{fig:lce_vs_reflectivity}
\end{figure}

As described in section~\ref{sec:geant4_optical}, most input parameters in the simulation framework can be determined from literature data or independent measurements, while the effective reflectivity $\rho$ for the Reflector configuration is the only key parameter that must be constrained by experiment. To this end, the peak position ratio between Grease and Air coupling under the same geometry is constructed:
\begin{equation}
    \mathcal{R}_{\mathrm{A-G}} \equiv \frac{V_{\text{peak}}(\text{grease})}{V_{\text{peak}}(\text{air})} .
\end{equation}
Since the Air and Grease configurations share the same electronics readout chain, the relative ratio $\mathcal{R}_{\mathrm{A-G}}$ of pulse amplitudes reflects purely the change in light collection efficiency due to the refractive index matching at the interface. On the simulation side, the equivalent ratio $\mathcal{R}_{\mathrm{A-G,sim}}(\rho) \equiv \alpha_{\mathrm{SiPIN}}(\mathrm{grease},\rho)\,/\,\alpha_{\mathrm{SiPIN}}(\mathrm{air},\rho)$ is constructed, and this ratio is found to be strictly monotonically decreasing with the cavity wall reflectivity $\rho$ (see figure~\ref{fig:lce_vs_reflectivity}).

The physical origin of this monotonic dependence lies in the competition between the modulation of the photon escape cone by the coupling medium and the effect of multiple diffuse reflections within the cavity. Under Air coupling ($n_{\mathrm{air}}=1.0$), total internal reflection at the crystal--air interface severely limits the solid angle of escapable photons, so that transmitted photons are biased toward small angles; whereas Grease coupling ($n_{\mathrm{grease}} \approx 1.46$) greatly alleviates the refractive index mismatch, allowing photons with a wider angular distribution to penetrate the interface, thereby significantly increasing the single-hit light collection probability. At the low reflectivity (low $\rho$) limit, photon collection depends mainly on the first direct transmission from crystal to detector, and the interface transmittance difference due to the coupling medium is maximised, leading to a high value of $\mathcal{R}_{\mathrm{A-G}}$; conversely, as the wall reflectivity $\rho$ increases, multiple diffuse reflections within the cavity promote full angle randomisation, and the local difference in single-pass interface transmittance is gradually smoothed by the complex traversal process, narrowing the light collection efficiency difference between the two paths and causing $\mathcal{R}_{\mathrm{A-G}}$ to approach~1 asymptotically.

The experimentally measured ratio for the Reflector configuration is $\mathcal{R}_{\mathrm{A-G,exp}} = 1.102 \pm 0.006$. Comparing this value with the simulation scan curve yields the effective reflectivity:
\begin{equation}
    \rho_{\text{eff}} = 0.92 \pm 0.01 .
    \label{eq:rho_eff}
\end{equation}
This result is lower than the reflectivity of ideal TiO$_2$ surfaces in the literature ($\sim 95.5\%$ \cite{janecekReflectivitySpectra2012}), consistent with the physical expectation of coating process limitations in a confined cavity and non-ideal surface losses. In subsequent discussion, the Reflector configuration simulation uses $\rho_{\text{eff}}=0.92$ as input.

\subsection{Light yield determination and consistency analysis}

Based on the $\rho_{\text{eff}}$ (Reflector) determined in the previous section and the manufacturer-specified $\rho=0.6\%$ (Absorber), $\alpha_{\text{SiPIN}}$ is computed for each of the four configurations and $LY_{\text{int}}$ is derived from equation~\eqref{eq:full_correction}. Table~\ref{tab:four_config_ly} summarises the measured and derived quantities for each configuration.

\begin{table}[htbp]
    \centering
    \caption{Summary of light yield measurement and correction for the four configurations (GAGG:Ce $5\times5\times5$~mm$^{3}$, S3590-08 SiPIN; $^{22}$Na 1274.5~keV photopeak, $N_{\text{rep}}=5$ per configuration)}
    \label{tab:four_config_ly}
    \begin{tabular}{lcccc}
        \toprule
        Configuration & $V_{\text{peak}}$\,(V) & $\alpha_{\text{SiPIN}}$\,(\%) & $Y_{ehp}$\,(ehp/MeV) & $LY_{\text{int}}$\,($\times 10^{3}$\,ph/MeV) \\
        \midrule
        Absorber--Air      & $0.3004\pm0.0022$ & 19.71 & 10\,985 & 55.7 \\
        Absorber--Grease   & $0.5062\pm0.0032$ & 31.98 & 18\,512 & 57.9 \\
        Reflector--Air     & $0.9300\pm0.0028$ & 60.86 & 34\,006 & 55.9 \\
        Reflector--Grease  & $1.0244\pm0.0042$ & 67.03 & 37\,458 & 55.9 \\
        \midrule
        \multicolumn{4}{r}{Mean $\pm$ spread $\pm$ syst} & $(56.3 \pm 1.0 \pm 1.6) \times 10^{3}$ \\
        \bottomrule
    \end{tabular}
\end{table}

\begin{figure}[htbp]
    \centering
    \includegraphics[width=\textwidth]{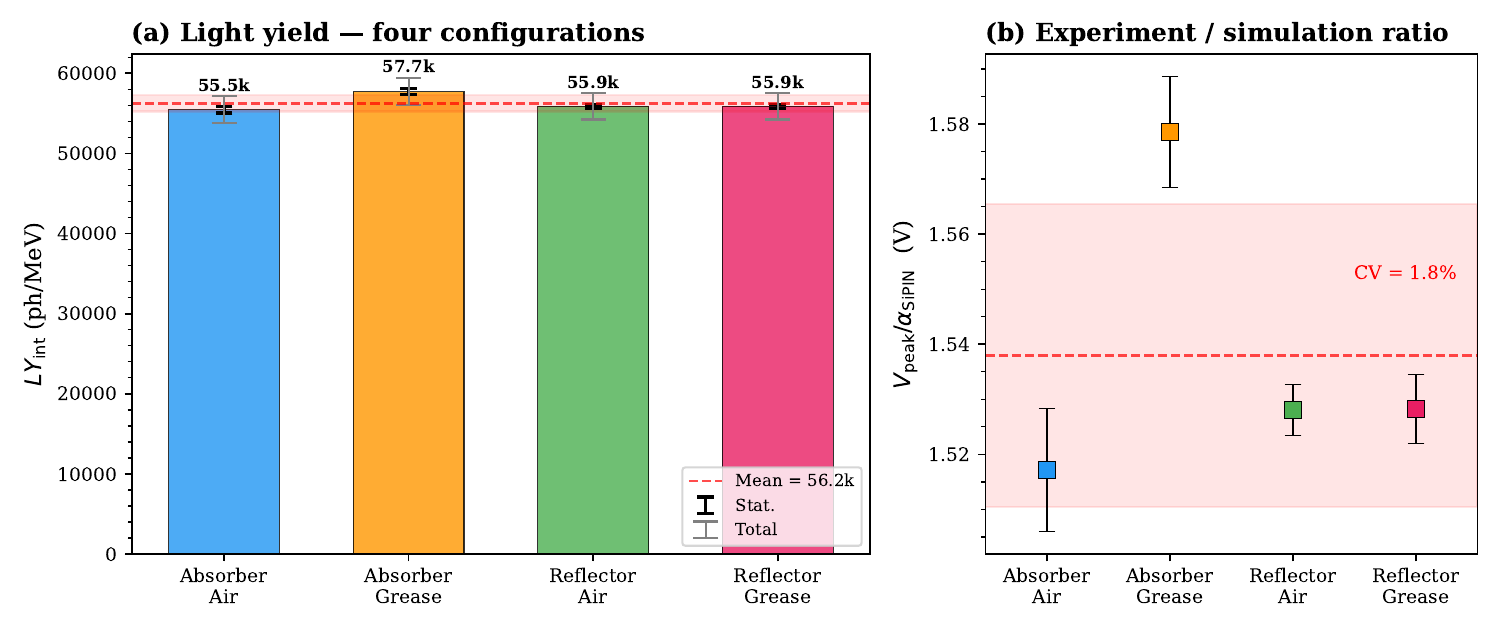}
    \caption{(a) Intrinsic light yield $LY_{\mathrm{int}}$ independently derived from four measurement configurations. Error bars: inner (black) = statistical; outer (gray) = total (including $\alpha_{\mathrm{SiPIN}}$ systematic). Dashed line: four-config mean; shaded band: $\pm 1\sigma$ spread. (b) $V_{\mathrm{peak}} / \alpha_{\mathrm{SiPIN}}$ ratio across configurations (coefficient of variation = 1.8\%).}
    \label{fig:light_yield_consistency}
\end{figure}

The $LY_{\text{int}}$ values derived from the four configurations show good consistency (figure~\ref{fig:light_yield_consistency}). The $LY_{\text{int}}$ deviation between Air and Grease within the Reflector group is only $0.2\%$, validating the self-consistency of the $\mathcal{R}_{\mathrm{A-G}}$ constraint method. More importantly, the comparison between the Absorber and Reflector paths: the key parameter $\rho$ for the two paths is obtained from completely independent sources (manufacturer specification vs.\ experimental constraint), and the resulting $\alpha_{\text{SiPIN}}$ spans a dynamic range of $20\%$--$67\%$ (a factor of $3.4$), yet the $LY_{\text{int}}$ deviation between the two paths is only $0.4\%$. Combining the four results, the overall coefficient of variation (CV) is $1.8\%$. Further analysis of this cross-configuration consistency for the reliability of the correction method is given in section~\ref{sec:model_validation}.

Combining the four results, the intrinsic light yield of the GAGG:Ce sample is determined to be:
\begin{equation}
    LY_{\text{int}} = (5.63 \pm 0.10_{\text{spread}} \pm 0.16_{\text{syst}})\times 10^{4}~\text{ph/MeV} \quad (\text{i.e.\ }56.3 \pm 1.9~\text{ph/keV}).
    \label{eq:LY_result}
\end{equation}
The first uncertainty is from the spread among the four configurations ($1030$~ph/MeV), and the second is propagated from the $\alpha_{\text{SiPIN}}$ systematic uncertainty ($\pm 2.9\%$, approximately $1634$~ph/MeV); the latter is the dominant contribution (see section~\ref{sec:uncertainties}). Additionally, in the commonly used Reflector--Grease configuration for scintillator measurements, the simulation gives $\alpha_{\text{SiPIN}} = 67.0\%$ ($\rho_{\text{eff}} = 0.92$, S3590-08), which can serve as a reference for evaluating SiPIN readout efficiency under similar geometric conditions.

\section{Discussion}
\label{sec:discussion}

This section provides further analysis of the measurement and correction results presented above: first, a sensitivity analysis identifies the dominant uncertainty sources of $\alpha_{\text{SiPIN}}$ and presents the uncertainty budget (section~\ref{sec:uncertainties}); then, the implications of cross-configuration consistency for the reliability of the correction method are discussed (section~\ref{sec:model_validation}); finally, the influence of different packaging boundary conditions on $\alpha_{\text{SiPIN}}$ is examined (section~\ref{sec:packaging_and_applicability}).

\subsection{Sensitivity analysis and systematic uncertainty}
\label{sec:uncertainties}

\begin{figure}[htbp]
    \centering
    \includegraphics[width=\textwidth]{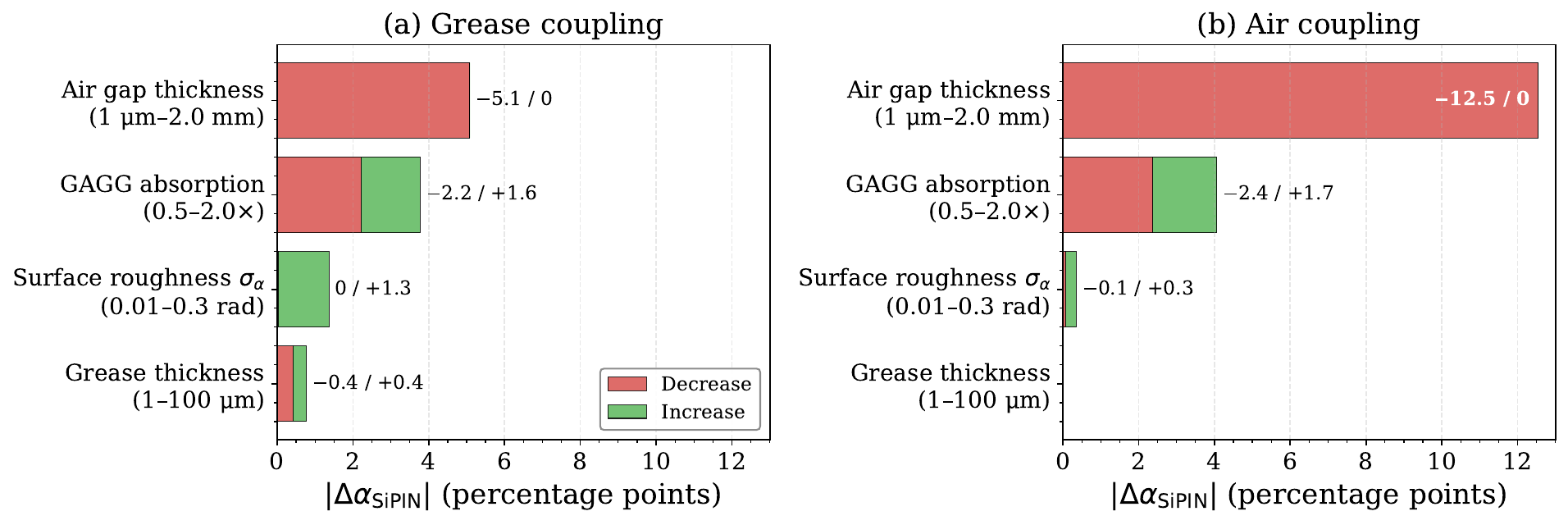}
\caption{Single-variable sensitivity analysis of $\alpha_{\text{SiPIN}}$ with respect to model parameters (tornado plot). Left: Grease coupling; right: Air coupling. Red segments indicate the decrease in $\alpha_{\text{SiPIN}}$ relative to the baseline configuration when parameters are biased toward the adverse end; green segments indicate the increase. The response of reflector wall reflectivity $\rho$ is shown separately in figure~\ref{fig:lce_vs_reflectivity} due to its different scale.}
    \label{fig:tornado_sensitivity}
\end{figure}

The total uncertainty of $LY_{\text{int}}$ comprises two components: the experimental contribution (statistical error of $Y_{ehp}$) and the simulation contribution (systematic error of $\alpha_{\text{SiPIN}}$). On the experimental side, repeated measurements indicate that the relative spread of the peak position $V_{\text{peak}}$ under the Reflector--Grease baseline configuration is approximately $0.9\%$, contributing only marginally to $LY_{\text{int}}$. The systematic uncertainty on the simulation side arises mainly from the model input parameters of $\alpha_{\text{SiPIN}}$ (reflectivity, crystal absorption length, surface roughness, coupling layer thickness, etc.). This section first identifies the most sensitive input parameters through single-variable sensitivity analysis (figure~\ref{fig:tornado_sensitivity}, table~\ref{tab:sensitivity_summary}), then assigns reference uncertainty ranges to each parameter and performs a systematic uncertainty evaluation (table~\ref{tab:uncertainty_budget}).

Table~\ref{tab:sensitivity_summary} gives the numerical summary corresponding to figure~\ref{fig:tornado_sensitivity} (Reflector configuration, grease coupling). The ``scan range'' in the table is used only for sensitivity diagnostics and does not represent the actual uncertainty range of each parameter in the present experiment.

\begin{table}[htbp]
    \centering
\caption{Single-variable sensitivity analysis results (Reflector configuration, grease coupling; for diagnosing the most sensitive inputs, not for uncertainty budget)}
    \label{tab:sensitivity_summary}
    \begin{tabular}{lcccc}
        \toprule
        Parameter & Scan range & Baseline & $\alpha_{\text{SiPIN}}$ range (\%) & Half-range \\
        \midrule
        Reflector wall reflectivity $\rho$ & 0.90--0.995 & 0.92 & 64.5--80.0 & $\pm 7.8$ \\
        GAGG intrinsic self-absorption ($d_{\text{path}}$) & 554--2215\,mm & 1108\,mm & 64.8--68.6 & $\pm 1.9$ \\
        Air gap thickness & 0.001--2.0\,mm & 2.0\,mm & 62.0--67.1 & $\pm 2.5$ \\
        Surface roughness $\sigma_{\text{surf}}$ & 0.01--0.3\,rad & 0.02\,rad & 67.0--68.4 & $\pm 0.7$ \\
        Grease thickness & 1--100\,$\mu$m & 50\,$\mu$m & 66.6--67.4 & $\pm 0.4$ \\
        \bottomrule
    \end{tabular}
\end{table}

The sensitivity analysis shows that reflector wall reflectivity $\rho$ has the strongest influence on $\alpha_{\text{SiPIN}}$. As shown in figure~\ref{fig:lce_vs_reflectivity}, $\alpha_{\text{SiPIN}}$ increases with $\rho$, varying by approximately $15.5$ percentage points over the range $\rho=0.90$--$0.995$. Near $\rho=0.92$, a $1\%$ decrease in $\rho$ leads to a decrease in $\alpha_{\text{SiPIN}}$ of approximately $1.3$ percentage points. The mechanism is as follows: in a non-tight-wrapping diffuse-reflector cavity, photons typically undergo multiple reflections before reaching the readout surface; when each reflection carries an absorption probability of $(1-\rho)$, the cumulative loss is significantly amplified as the mean number of reflections increases (estimable on the order of $\rho^{N_{\text{bounce}}}$).

$\rho$ also determines the simulated Grease/Air ratio $\mathcal{R}_{\mathrm{A-G,sim}}$. The effective reflectivity $\rho_{\text{eff}} = 0.92 \pm 0.01$ is constrained from the experimental value $\mathcal{R}_{\mathrm{A-G,exp}}=1.102\pm0.006$ (equation~\eqref{eq:rho_eff}). Taking $\rho_{\text{eff}} \approx 0.92$ as the representative value, the range $0.91$--$0.93$ is adopted as the reference uncertainty range (corresponding to $\Delta\rho=\pm0.01$ giving $\Delta\alpha\approx\pm1.4$ percentage points).

\begin{figure}[htbp]
    \centering
    \includegraphics[width=\textwidth]{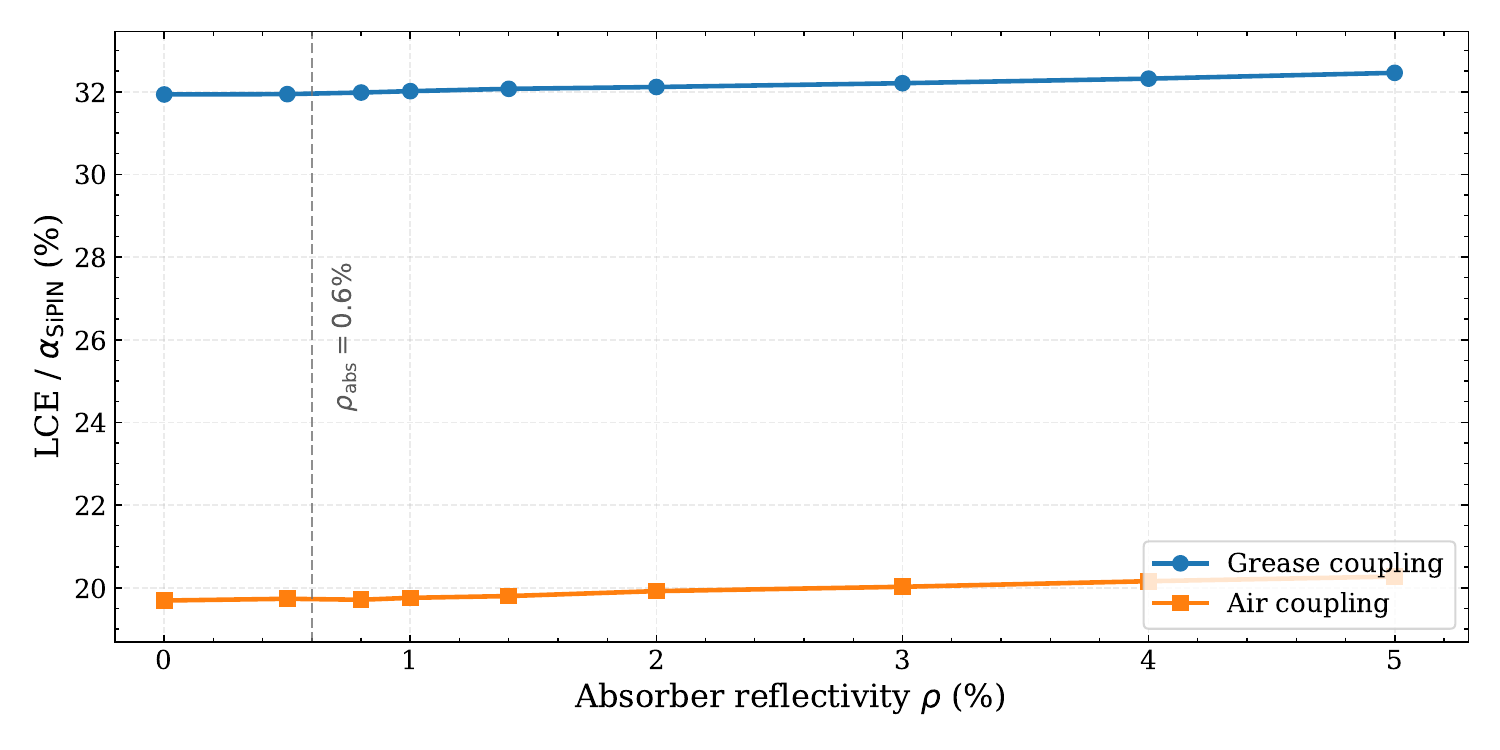}
    \caption{Variation of $\alpha_{\text{SiPIN}}$ with housing reflectivity $\rho$ under the Absorber configuration ($\rho = 0$--$5\%$). The dashed line marks $\rho_{\text{abs}} = 0.6\%$ used in the present experiment. In contrast to the strong dependence in the Reflector configuration shown in figure~\ref{fig:lce_vs_reflectivity}, $\alpha_{\text{SiPIN}}$ is nearly insensitive to $\rho$ in the Absorber configuration.}
    \label{fig:lce_vs_low_reflectivity}
\end{figure}

In contrast, the response under the Absorber configuration ($\rho < 5\%$) is markedly different. Figure~\ref{fig:lce_vs_low_reflectivity} shows the $\alpha_{\text{SiPIN}}$ scan results for $\rho$ in the range $0$--$5\%$: for Grease coupling, $\alpha_{\text{SiPIN}}$ varies only between $31.9\%$ and $32.5\%$ (total span $0.5$ percentage points), and for Air coupling between $19.7\%$ and $20.3\%$ (span $0.6$ percentage points). Near $\rho_{\text{abs}}\approx 0.6\%$, a $1\%$ change in $\rho$ induces only about $0.1$--$0.2$ percentage points shift in $\alpha_{\text{SiPIN}}$, one order of magnitude lower than the sensitivity in the Reflector configuration. This insensitivity arises because in the Absorber cavity, photons reach the SiPIN mainly via direct paths, with multiple reflections contributing negligibly; changes in $\rho$ affect only the small fraction of indirect photons that reach the SiPIN after reflection from the housing, and thus have negligible impact on the total $\alpha_{\text{SiPIN}}$. This conclusion also corroborates the robustness of the Absorber configuration for $\alpha_{\text{SiPIN}}$ correction: even with substantial calibration uncertainty in the housing reflectivity, the impact on the inferred light yield does not exceed $0.5\%$. Therefore, variations in coating technique and process can be neglected for $\alpha_{\text{SiPIN}}$, and the supplier-specified $\rho \approx 0.6\%$ can be used directly as the simulation input.

\begin{figure}[htbp]
    \centering
    \includegraphics[width=\textwidth]{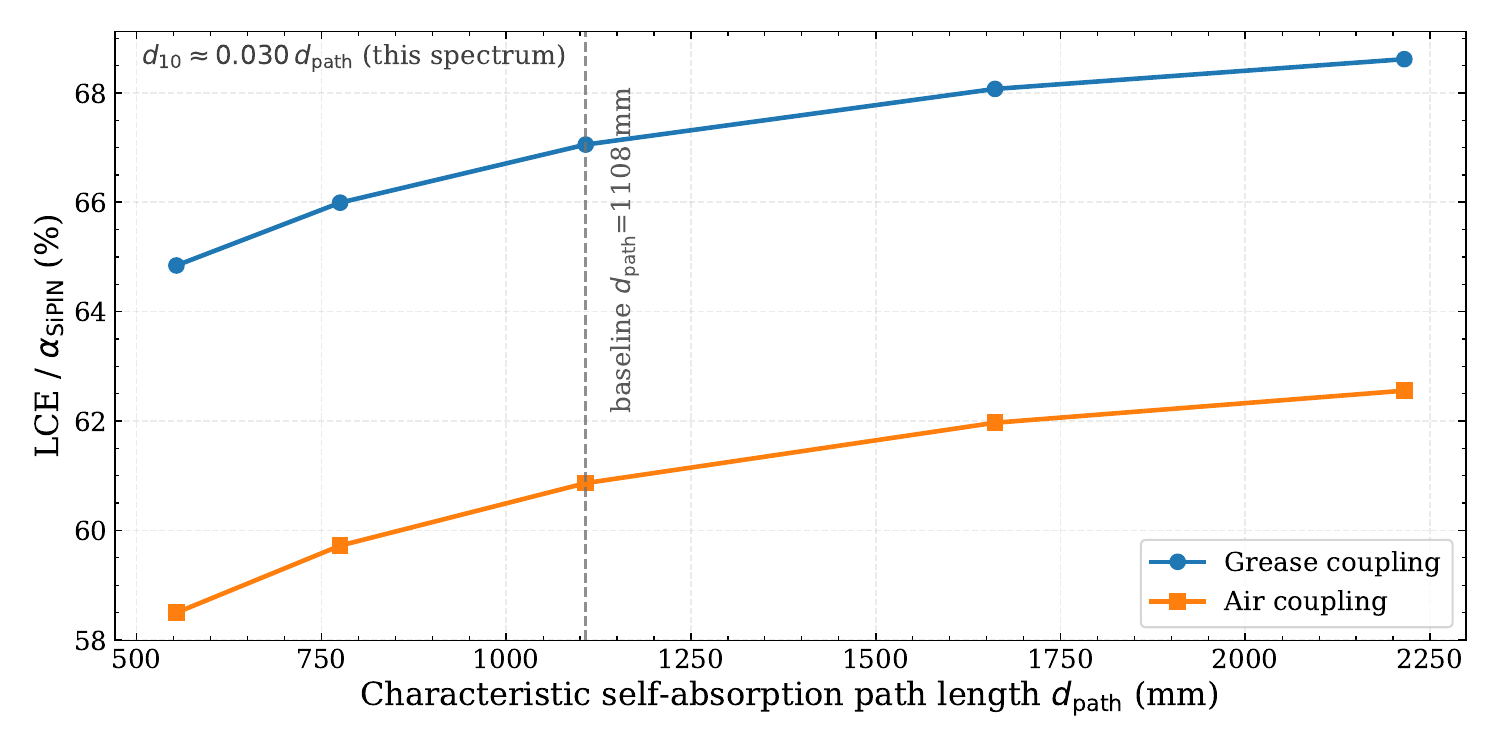}
\caption{Variation of $\alpha_{\text{SiPIN}}$ with the GAGG intrinsic self-absorption characteristic optical path $d_{\text{path}}$. Smaller $d_{\text{path}}$ corresponds to stronger self-absorption.}
    \label{fig:lce_vs_absorption}
\end{figure}

The second most influential factor is crystal self-absorption. Owing to the large Stokes shift of GAGG:Ce, the overlap between its emission and absorption spectra is small, making it a weakly self-absorbing material. For convenience, the wavelength-dependent absorption length $L_{\text{abs}}(\lambda)$ is characterised by a single parameter, and the spectrally weighted self-absorption fraction is defined as
\begin{equation}
    J(\ell)=\int F(\lambda)\left[1-\exp\left(-\frac{\ell}{L_{\mathrm{abs}}(\lambda)}\right)\right]\mathrm{d}\lambda,
    \label{eq:self_absorption_overlap}
\end{equation}
where $F(\lambda)$ is the normalised emission spectrum. The characteristic optical path $d_{\text{path}}$ (also denoted $d_{1/e}$) is further defined by the $1/e$ convention (satisfying $J(d_{\text{path}})=1-1/e$). Figure~\ref{fig:lce_vs_absorption} shows that even when $d_{\text{path}}$ varies over a wide range of $554$--$2215$\,mm (baseline $d_{\text{path}}=1108$\,mm, covering from typical to high-transparency samples), the change in $\alpha_{\text{SiPIN}}$ is only about $\pm 1.9$ percentage points. This indicates that for small GAGG samples, the correction result is insensitive to measurement errors in the absorption coefficient.

Table~\ref{tab:uncertainty_budget} summarises the systematic uncertainty of $\alpha_{\text{SiPIN}}$ under the Reflector configuration (Grease coupling). The dominant term is reflector wall reflectivity $\rho$, which is constrained to $\rho_{\text{eff}} = 0.92 \pm 0.01$ via the experimental ratio $\mathcal{R}_{\mathrm{A-G}}$, corresponding to $\delta\alpha/\alpha \approx \pm 2.0\%$. The contributions from the remaining optical parameters (absorption length, air gap thickness, surface roughness, coupling layer thickness) do not exceed $1.5\%$. The model uncertainty of the TMM equivalent model is evaluated to be approximately $0.37\%$ (see section~\ref{sec:tmm}), conservatively taken as $\pm 0.4\%$. Combining the upper limits of each term in quadrature, the combined systematic uncertainty of $\alpha_{\text{SiPIN}}$ is approximately $\pm 2.9\%$, lower than the typical $\pm 5\%$ level in conventional PMT-based relative-method absolute light yield measurements \cite{moszynskiAbsoluteLightOutput1997}.

For the propagation of uncertainty to $LY_{\text{int}}$, the first-order approximation of the relative uncertainty from equation~\eqref{eq:full_correction} is:
\begin{equation}
    \left(\frac{\sigma_{LY_{\text{int}}}}{LY_{\text{int}}}\right)^2
    \simeq
    \left(\frac{\sigma_{Y_{ehp}}}{Y_{ehp}}\right)^2
    +
    \left(\frac{\sigma_{\alpha_{\text{SiPIN}}}}{\alpha_{\text{SiPIN}}}\right)^2 .
\end{equation}
Here $\sigma_{Y_{ehp}}$ arises from the experimental uncertainty of $Y_{ehp}$, and $\sigma_{\alpha_{\text{SiPIN}}}$ from the systematic uncertainty of $\alpha_{\text{SiPIN}}$. Comparing the experimental statistical error ($\sim 0.9\%$) with the simulation systematic error ($\sim 2.9\%$, table~\ref{tab:uncertainty_budget}), the total uncertainty of $LY_{\text{int}}$ is dominated by the systematic contribution of $\alpha_{\text{SiPIN}}$. In addition, all measurements are based on a single energy point (the $1274.5$~keV photopeak of $^{22}$Na), so no non-proportionality correction is required.

\begin{table}[htbp]
    \centering
    \caption{Systematic uncertainty budget for $\alpha_{\text{SiPIN}}$ (Reflector, Grease coupling).}
    \label{tab:uncertainty_budget}
    \small
    \begin{tabular}{lccc}
        \toprule
        Source & Range & $\delta\alpha/\alpha$ & Constraint \\
        \midrule
        Wall reflectivity $\rho$ & $0.92\pm0.01$ & $\pm 2.0\%$ & Exp.\ $\mathcal{R}_{\mathrm{A-G}}$ \\
        GAGG absorption ($d_{\text{path}}$) & $\pm 30\%$ & $\pm 1.5\%$ & Weak self-absorption \\
        Air gap thickness & 1--3\,mm & ${<}\,1\%$ & Geometric tolerance \\
        Roughness $\sigma_{\text{surf}}$ & 0.01--0.1\,rad & ${<}\,0.5\%$ & Polished; TIR-dominated \\
        Grease thickness & 10--100\,$\mu$m & ${<}\,0.5\%$ & Index matching \\
        TMM model & AR + linear $T_{\text{epoxy}}$ & $\pm 0.4\%$ & See section~\ref{sec:tmm} \\
        \midrule
        Combined & \multicolumn{2}{c}{$\pm 2.9\%$ (quadrature sum)} & \\
        \bottomrule
    \end{tabular}
\end{table}

\subsection{Validation of the correction method}
\label{sec:model_validation}

The $1.8\%$ level consistency among the four experimentally measured configurations validates the reliability of the correction method from two perspectives: internal self-consistency and cross-validation across boundary conditions. First, within the Reflector path, the calibration of effective reflectivity $\rho_{\text{eff}}$ relies only on the relative ratio $\mathcal{R}_{\mathrm{A-G}}$ under different coupling media, thereby decoupling potential systematic bias from absolute electronics gain. Under this constraint, the relative deviation in $LY_{\text{int}}$ derived from Air and Grease coupling within the Reflector group is only $0.2\%$, confirming the self-consistency of the model for a single reflector geometry.

Second, the cross-comparison between the Absorber and Reflector configurations provides an independent and stringent test of the model. The two optical boundary conditions exhibit markedly different sensitivity distributions to model parameters: the $\alpha_{\text{SiPIN}}$ of the Absorber group ($20\%$--$32\%$, $\rho=0.6\%$) is dominated by direct photons from the crystal to the SiPIN, and is therefore sensitive to the values of $p_{\text{det}}(\lambda,\theta)$ in the low-to-mid incidence angle range; whereas the $\alpha_{\text{SiPIN}}$ of the Reflector group ($61\%$--$67\%$) depends strongly on multiple diffuse reflections, and its detection efficiency is more sensitive to $\rho_{\text{eff}}$ and $p_{\text{det}}$ in the large-angle regime. Since no single parametric systematic bias can simultaneously compensate for these two readout paths with orthogonal sensitivity characteristics (e.g., a bias in large-angle $p_{\text{det}}$ would significantly affect Reflector but have almost no effect on Absorber), the agreement in light yield derived from these two independently parameterised configurations demonstrates the accuracy of the simulation framework in handling complex angular distributions and multi-interface reflection transport.

Finally, the analysis of the physical origin of the cross-group deviation further supports the reasonableness of the model. In the Absorber cavity, light collection is dominated by direct paths, lacking the angular averaging effect of multiple diffuse reflections, so the system is more sensitive to crystal position deviation and local coupling state; in the Reflector cavity, multiple diffuse reflections effectively smooth out the effects of geometric asymmetry. The experimental result that the Absorber--Grease group yields a light yield $+2.7\%$ higher is consistent with the physical expectation that local assembly fluctuations are amplified under low-reflectivity boundaries, while the Reflector light collection scheme exhibits better measurement repeatability.

\subsection{Influence of packaging boundary conditions}
\label{sec:packaging_and_applicability}

To assess the applicability of the method under different packaging conditions, this section compares the behaviour of $\alpha_{\text{SiPIN}}$ under two typical boundary conditions: Reflector (reflective housing with air gap) and Teflon wrapping (tight wrapping without air gap).

In the Reflector configuration (the experimental baseline of this work), there is a millimetre-scale air gap between the crystal and the reflector layer. The air gap allows the crystal side and top surfaces to satisfy the TIR condition, so photons can propagate within the crystal with low interface loss through multiple reflections; only photons falling into the escape cone exit the crystal and interact with the reflector. This boundary condition explains why surface roughness $\sigma_{\text{surf}}$ has a small effect on $\alpha_{\text{SiPIN}}$ in this configuration (see table~\ref{tab:sensitivity_summary}, variation ${<}0.5\%$): when TIR dominates, the frequency of photon contact with the lossy reflector is reduced, and the effect of roughness on angular randomisation is correspondingly diminished.

\begin{figure}[htbp]
    \centering
    \includegraphics[width=\textwidth]{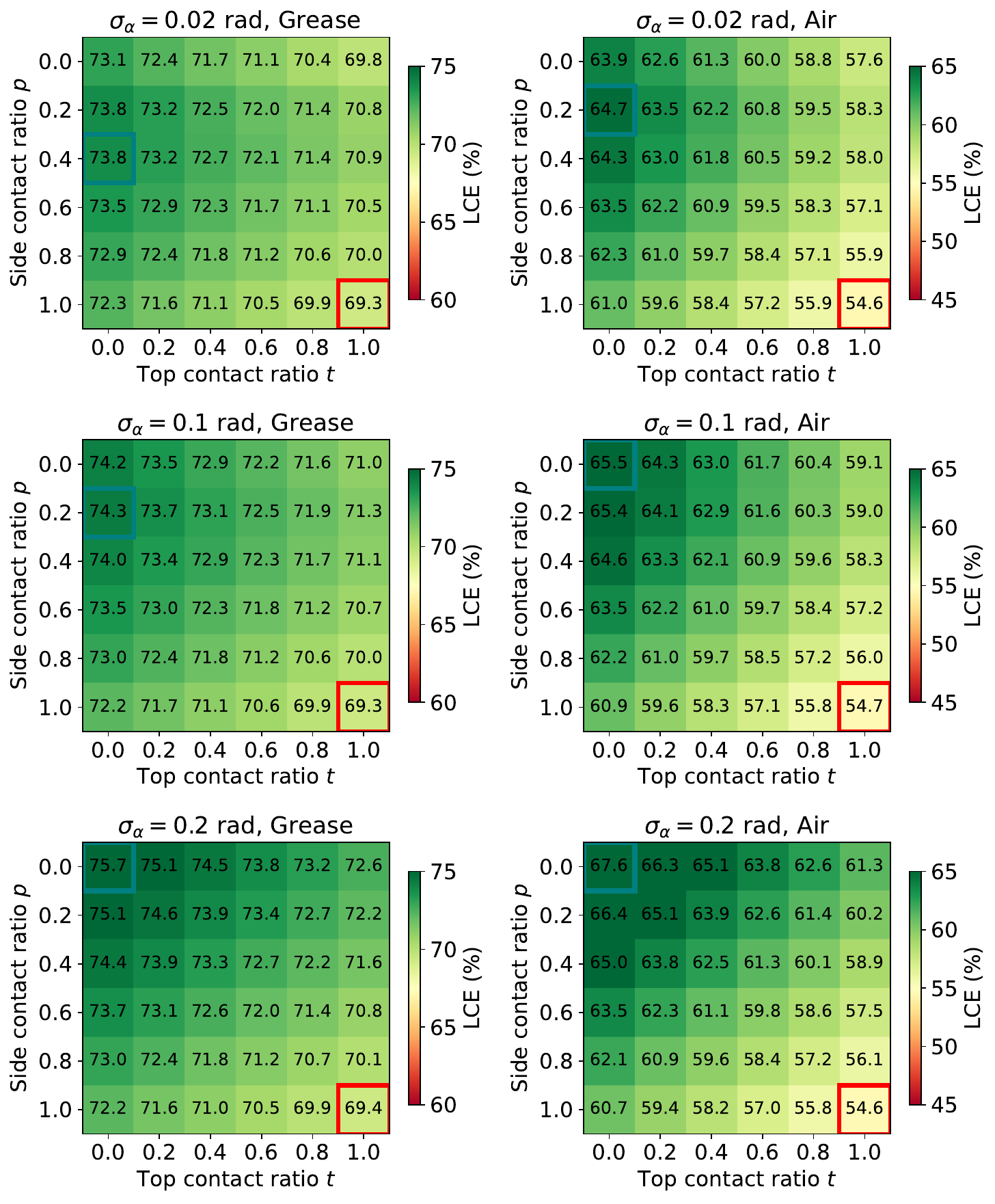}
    \caption{Heat map of $\alpha_{\text{SiPIN}}$ versus side contact fraction $p$, top contact fraction $t$, and surface roughness $\sigma_{\text{surf}}$ under Teflon tight wrapping (PTFE reflectivity 97.5\%, air gap = 1\,$\mu$m). Colour scale corresponds to $\alpha_{\text{SiPIN}}$ (\%). Green and red boxes mark the best and worst contact conditions, respectively, for each case.}
    \label{fig:teflon_lce_heatmap_sigma}
\end{figure}

In contrast, Teflon tight wrapping (air gap $\sim 1\,\mu$m) brings the crystal surface and reflector into the near-field coupling regime, partially invalidating the angular range originally protected by TIR. To approximate the packaging state where ``local contact and local air layer coexist'', the contact fraction $(p,t)$ is introduced to represent the area fraction of the crystal side and top surfaces in direct contact with the reflector. In the simulation, each time a photon hits the side (top) surface, a random draw with probability $p$ (or $t$) determines whether the interaction occurs via direct contact: if so, it is treated with the PTFE (polytetrafluoroethylene, i.e.\ Teflon) reflectivity and Lambertian diffuse reflection model; otherwise, Geant4 handles it as a crystal--air interface Fresnel process (with TIR when the critical angle condition is met). In this scan, crystal surface roughness is implemented via the UNIFIED model parameter $\sigma_{\text{surf}}$, whose effect is mainly on the non-contacting crystal--air-gap boundary, characterising the angular randomisation from micro-facet normal broadening. Figure~\ref{fig:teflon_lce_heatmap_sigma} shows the dependence of $\alpha_{\text{SiPIN}}$ on contact fraction $(p, t)$ and $\sigma_{\text{surf}}$ under the above model; this scan serves to quantify the magnitude and mechanism of packaging boundary effects on $\alpha_{\text{SiPIN}}$, and is not part of the main measurement conclusions of this work.

\begin{table}[htbp]
    \centering
    \caption{Summary of $\alpha_{\text{SiPIN}}$ for Teflon tight wrapping (different $\sigma_{\text{surf}}$, PTFE reflectivity 97.5\%, air gap = 1\,$\mu$m)}
    \label{tab:teflon_sigma_summary}
    \begin{tabular}{lccccc}
        \toprule
        $\sigma_{\text{surf}}$ (rad) & Coupling & $\alpha_{\text{SiPIN}}$ range (\%) & Best $(p, t)$ & Worst $(p, t)$ & Max.\ variation \\
        \midrule
        0.02 & Grease & 69.3--73.8 & (0.4, 0.0) & (1.0, 1.0) & 4.5\% \\
        0.02 & Air & 54.6--64.7 & (0.2, 0.0) & (1.0, 1.0) & 10.1\% \\
        0.1 & Grease & 69.3--74.3 & (0.2, 0.0) & (1.0, 1.0) & 5.0\% \\
        0.1 & Air & 54.7--65.5 & (0.0, 0.0) & (1.0, 1.0) & 10.8\% \\
        0.2 & Grease & 69.4--75.7 & (0.0, 0.0) & (1.0, 1.0) & 6.3\% \\
        0.2 & Air & 54.6--67.6 & (0.0, 0.0) & (1.0, 1.0) & 12.9\% \\
        \bottomrule
    \end{tabular}
\end{table}

The following patterns can be observed from figure~\ref{fig:teflon_lce_heatmap_sigma}. For all $\sigma_{\text{surf}}$, the worst point occurs at $(p, t) = (1, 1)$, i.e., full contact on both side and top surfaces. This trend can be understood in terms of the cumulative effect of lossy reflection: when all surfaces are in direct contact with PTFE, every interface reflection of a photon is accompanied by $(1-\rho)$ absorption loss, TIR is effectively disabled, and losses accumulate significantly over multiple reflections. In contrast, the best point at low $\sigma_{\text{surf}}$ ($\leq 0.1$\,rad) lies near $(p, t) \approx (0.2\text{--}0.4,\, 0)$, indicating that moderate side contact helps to alter the angular distribution via diffuse reflection and increase the probability of reaching the readout surface, while keeping an air gap on the top preserves TIR as a low-loss ``mirror''; when $\sigma_{\text{surf}}$ increases to $0.2$\,rad, the best point shifts to $(0,\,0)$, i.e., no contact at all---in this case, micro-facet scattering at the crystal surface already provides sufficient angular mixing, and additional PTFE contact only adds absorption loss.

The results show that as $\sigma_{\text{surf}}$ increases from $0.02$\,rad to $0.2$\,rad, the sensitivity of $\alpha_{\text{SiPIN}}$ to contact state actually increases: the maximum variation rises from $4.5\%$ to $6.3\%$ for grease coupling and from $10.1\%$ to $12.9\%$ for air coupling. The mechanism is that larger $\sigma_{\text{surf}}$ significantly raises $\alpha_{\text{SiPIN}}$ in the non-contacting region (retaining air gap)---e.g., the $(0,0)$ point in the air case increases from $64.7\%$ to $67.6\%$---while $\alpha_{\text{SiPIN}}$ at the full-contact $(1,1)$ point hardly changes with $\sigma_{\text{surf}}$ (remaining at $\sim 54.6\%$), thereby widening the gap between the best and worst points. Physically, larger $\sigma_{\text{surf}}$ broadens the micro-facet normal distribution, providing more effective angular randomisation at the air-gap-retaining interface, so that more photons are scattered from the escape cone back into the crystal and eventually reach the readout surface; at the fully contacting interface, regardless of $\sigma_{\text{surf}}$, photons always interact directly with PTFE, and roughness no longer plays a protective role.

The packaging scan in this section aims to reveal the magnitude and main mechanisms of $\alpha_{\text{SiPIN}}$ variation under different boundary conditions, and to avoid directly extrapolating the calibration and validation results from the baseline configuration to all packaging geometries. For Reflector-type packaging with an air gap, the uncertainty of $\alpha_{\text{SiPIN}}$ is dominated by reflectivity $\rho$ and air gap geometric tolerance; for tight-wrapping packaging, there is strong coupling between contact state and surface roughness, and the associated uncertainty is harder to characterise with a single parameter. Therefore, when pursuing controllable and reproducible absolute measurements, a reflective housing structure with geometric constraints and a well-defined air gap is generally more favourable for obtaining stable, traceable assembly conditions.

\section{Conclusions}
\label{sec:conclusions}

This paper establishes a SiPIN-based absolute light yield calibration method that combines the transfer-matrix method (TMM) with Geant4 simulation. The method uses TMM to reconstruct the manufacturer's standard quantum efficiency (QE) into an angle- and wavelength-dependent single-hit detection probability $p_{\mathrm{det}}(\lambda,\theta)$, which is then dynamically integrated into the Geant4 optical transport. This framework treats in a physically self-consistent way the effects of wide-angle incidence, multiple interface hits, and coupling-medium changes on detector response within the macroscopic transport simulation.

Systematic experimental validation was carried out using a $5\times5\times5~\mathrm{mm}^3$ GAGG:Ce crystal. Based on an absolute charge calibration reference from a $^{241}$Am source, cross-checks were performed along two optically distinct paths with fully orthogonal parameter sources: low reflectivity (Absorber, $\rho \approx 0.6\%$) and high reflectivity (Reflector, constrained in situ to $\rho_{\mathrm{eff}}=0.92\pm0.01$). Although the $\alpha_{\mathrm{SiPIN}}$ values of the four configurations span a large dynamic range from $20\%$ to $67\%$, the intrinsic light yields derived from each independent configuration show excellent cross-boundary consistency (coefficient of variation of only $1.8\%$). The intrinsic light yield of GAGG:Ce is determined to be $LY_{\mathrm{int}} = (5.63 \pm 0.10_{\mathrm{spread}} \pm 0.16_{\mathrm{syst}})\times 10^{4}~\mathrm{ph/MeV}$ (i.e.\ $56.3 \pm 1.9~\mathrm{ph/keV}$). The systematic uncertainty evaluation indicates that the current total systematic uncertainty of $\pm 2.9\%$ is mainly limited by the precision with which the effective reflectance of the high-reflectance cavity is constrained, while the model shows good robustness to local geometric parameters such as coupling layer thickness and surface roughness.

The calibration method established in this paper has well-defined applicability conditions: the scintillator emission spectrum and bulk absorption must be independently characterisable, the detector window response must have external data constraints, and the measurement setup must have reproducible macroscopic assembly. For tight-wrapping systems (e.g.\ Teflon), the strong coupling between interface contact fraction and microscopic surface roughness can easily introduce systematic bias that is difficult to parametrise with a single quantity; practical applicability in such cases remains to be further explored. In principle, the method is applicable to other scintillator materials (e.g.\ LYSO:Ce, LaBr$_3$:Ce) and silicon-based photodetector systems. In particular, for SiPM array readout, where the photon detection efficiency also has strong angular dependence, the TMM layer model can be replaced by an equivalent micro-cell response model for the SiPM, and the Geant4 geometry transport and experimental constraint strategy established in this paper can be reused. When higher absolute precision is required, independent constraints on reflector reflectivity and interface state, as well as in-situ characterisation, remain the main limiting factors.

\acknowledgments
This work is partly supported by the National Key Research and Development Program of China under grant No.\ 2022YFA1601900, National Natural Science Foundation of China (NSFC) under grant Nos.\ 11925504 and W2443007, and Peking University Faculty Start-up Fund.

\bibliographystyle{JHEP}
\bibliography{references}
\end{document}